\documentclass{aa}

\usepackage{graphicx}
\usepackage{txfonts}

\usepackage{soul}

\newcommand{\Fermi}{\emph{Fermi}\,}
\newcommand{\angstrom}{\textup{\AA}}

\usepackage{color}
\newcommand{\rd}{\color{black}}

\begin{document}

\title{Evidence for flare-accelerated particles in large scale loops in the behind-the-limb gamma-ray solar flare of September 29, 2022}

\author{Melissa Pesce-Rollins \inst{1}, Karl-Ludwig Klein \inst{2}, S\"am Krucker \inst{3,4}, Alexander Warmuth \inst{5}, Astrid M. Veronig \inst{6} , Nicola Omodei \inst{7} and Christian Monstein \inst{8}}

 \institute{Istituto Nazionale di Fisica Nucleare, Sezione di Pisa I-56127 Pisa, Italy 
 \and
 Observatoire de Paris, LESIA \& Observatoire Radio Astronomique de Nan\c{c}ay, Univ. PSL, CNRS, Sorbonne Univ., Univ. Paris Cit\'e, OSUC, 5 place Jules Janssen, F-92190 Meudon, France 
 \and
 University of Applied Sciences and Arts Northwestern Switzerland CH-5210 Windisch, Switzerland 
 \and Space Science Laboratory, University of California, Berkeley, CA 94720-7450, USA
\and
Leibniz-Institut f\"ur Astrophysik Potsdam (AIP) An der Sternwarte 16
14482 Potsdam, Germany
\and
Institute of Physics \& Kanzelh\"ohe Observarory, University of Graz,
Universit\"atsplatz 5, 8010 Graz, Austria
\and
W. W. Hansen Experimental Physics Laboratory, Kavli Institute for Particle Astrophysics and Cosmology, Department of Physics and SLAC National Accelerator Laboratory, Stanford University, Stanford, CA 94305, USA
 \and
 Istituto ricerche solari Aldo e Cele Dacc\`o (IRSOL), Faculty of Informatics, Universit\`a della Svizzera italiana (USI), CH-6605 Locarno, Switzerland.
 }

\authorrunning{Pesce-Rollins \& Klein et al.}
\titlerunning{Gamma-ray behind-the-limb flare of September 29, 2022}

\abstract
{We report on the detection of the gamma-ray emission above 100 MeV from the solar flare of September 29, 2022, by
\Fermi LAT with simultaneous coverage in HXR by Solar Orbiter STIX. The Solar Orbiter-Earth separation was 178$^{\circ}$ at the time of the flare as seen from Earth, with Solar Orbiter
observing the east limb. Based on STIX imaging, the flare was located 16$^{\circ}$ behind the eastern limb as seen from Earth. The STIX and GBM non-thermal emission and the LAT emission above 100 MeV  all show similarly shaped time profiles, and the \Fermi\ profiles peaked only 20 seconds after the STIX signal from the main flare site, setting this flare apart from all the other occulted flares observed by \Fermi LAT. The radio
spectral imaging based on the Nan\c{c}ay Radioheliograph and ORFEES spectrograph reveal geometries
consistent with a magnetic structure that connects the parent active region behind the limb to the
visible disk. We studied the basic characteristics of the gamma-ray time profile, in particular, the rise and decay times and the time delay between the gamma-ray and HXR peak fluxes. We compared the characteristics of this event with those of four \Fermi LAT behind-the-limb flares and with an on-disk event and found that this event is strikingly similar to the {\rd impulsive} on-disk flare. Based on multiwavelength observations, we find that the gamma-ray emission above 100 MeV originated from ions accelerated in the parent active region behind the limb and was transported to the visible disk via a large magnetic structure connected to the parent active region behind the limb. {\rd Our results strongly suggest that the source of the emission above 100 MeV from the September 29, 2022 flare cannot be the CME-driven shock.}}

\keywords{Sun: X-rays, gamma rays, microwave, radio, EUV -- Sun: flares }

\maketitle

\section{Introduction} \label{sec:intro}
{\rd Solar flares have been known to be sources of high-energy gamma rays for over 40 years now. Evidence that $\sim$3-50 MeV ions are being accelerated during the impulsive phase of solar flares was found by observations of nuclear de-excitation lines in the spectra~\citep{1982ApJ...263L..95C,1985ICRC....4..146F,FORREST1986115, 1994ApJ...425L.109B,
Ackermann_2012}, and in some bright flares, a greater than 100-MeV gamma-ray continuum has also been observed 
\citep{Aki:al-92,Vil:al-03,Chupp_2009,Msn:al-09}, 
indicating that ions were being accelerated to energies greater than 300 MeV. It has been generally accepted that the magnetic energy released through reconnection during solar flares was the mechanism responsible for accelerating ions to such high energies~\citep[e.g.][]{Shih_2009}.

While observations had already suggested that there were multiple phases in the gamma-ray solar flares~\citep{FORREST1986115}, the detection of gamma-ray emission persisting for hours after all other counterpart radiation had ceased~\citep{Kanbach1993, Chupp_2009,Ajello_2014,2014ApJ...787...15A} presented a challenge to the classical magnetic reconnection theory for ion acceleration. Flares with this type of phase have been called long-duration gamma-ray flares (LDGRFs) or sustained gamma-ray emission (SGRE), and several scenarios to explain the phenomenon have been proposed over the years. Two of the most popular are (a) acceleration at the coronal mass ejection (CME)-driven shock with back precipitation to the solar atmosphere~\citep{rank2001} and (b) trapping of flare-accelerated ions in extended coronal loops {or additional acceleration and release into the loop}~\citep{ryanlee91,1992ApJ...396L.111M,Lit:Som-95,ryan00,Her:al-02}.}

\Fermi\ Large Area Telescope (LAT)~\citep{LATPaper} observations of high-energy gamma-ray emission from solar flares over the past 15 years have now yielded a sample that is large enough for population studies~\citep{flarecatalog_2021, Share_2018}. A special subsample of these flares are those whose active region is located behind the visible limb at the time of the eruption, the so-called behind-the-limb (BTL) flares.  Since gamma-ray emission must come from relatively dense plasmas, observations
of these rare events pose interesting questions regarding the acceleration mechanisms at work and the transport of the accelerated particles. Prior to the launch of the \Fermi\ satellite, only three of these BTL gamma-ray flares had been observed to have emission with energies up to 100 MeV. The total number has now\footnote{At the time of writing this manuscript.} increased to nine thanks to the LAT detection of six BTL flares with observed energies in the gigaelectron volt range. 

{\rd The scenarios put forth to explain the emission of BTL flares are essentially the same as those for 
sustained gamma-ray emission. 
In order for the accelerated ions to reach the visible side of the disk, an extended source is invoked, 
and such a source could be the CME shock~\citep{cliv93, 1993ApJ...409L..69V, 
Ackermann_2017, Plotnikov_2017, Jin2018, 2020SoPh..295...18G,wu2021,Grechnev_2018}. 
}
{\rd Observational {results indicating a 
correlation between the durations of interplanetary type II radio bursts and gamma-ray flares}
as well as support via magnetohydrodynamic simulations and modeling reported in the literature \citep[e.g.][]{Jin2018, Plotnikov_2017,2018ApJ...868L..19G,2020SoPh..295...18G} have made} the CME-driven shock one of the most popular scenarios for explaining the gamma-ray emission from {\rd not only} these events {\rd but also for the flares with SGRE}. Additional evidence {\rd in support of the CME-shock scenario} has also been reported by  \cite{Pesce-Rollins_2022}, showing that there exists a coupling between the extreme ultraviolet (EUV) coronal wave and the ion acceleration occurring in four of the BTL flares detected by LAT. 
Such results are in line with the long tradition that considers the presence of a type II radio burst at meter waves and beyond as an indication of a shock wave capable of accelerating particles in general 
\citep{Boi:Den-57,Wil:al-63,Rea-09,Gop:al-12}.

The shock-acceleration scenario, however, is not without problems. The timing of gamma-ray emission from relativistic protons or electrons in the impulsive flare phase requires an extremely rapid acceleration that is unlikely to be possible at an extended quasi-parallel coronal shock \citep{Afa:al-18b}. The idea that relativistic particles from a CME shock could sustain gamma-ray emission over many hours requires a nearly unimpeded travel from large heliocentric distances back to the Sun despite magnetic mirroring \citep{Kle:al-18,Hud-18}. 
{\cite{DeNolfo2019} and \cite{Bruno_2023} analyzed the interacting and solar energetic particle ion populations during LDGRF events and found that very large precipitation fractions are required in several of the events in order for the CME-shock scenario to explain the observations. However, models of back precipitation\citep{Hutchinson2022} do not support large precipitation fractions, thus posing a problem for the CME scenario.} An alternative interpretation {to the CME-shock scenario of BTL gamma-ray events} is particle acceleration in the flaring active region and their injection into large coronal loops that connect to the disk \citep{Kle:al-99a,1999A&A...342..575V,Grechnev_2018, ryan00}. Though more observational evidence is needed, the BTL flares represent a test bed for this problem.

The first five $>$100 MeV BTL flares {\rd observed by LAT} have time profiles that resemble delayed emission, {\rd namely,} the time profiles have a slow rise and a slower decay and a flux peaking several minutes after the associated hard X-ray (HXR) flux. However, the detection of gamma rays from the BTL flare of September 29, 2022, adds a new twist to the story because this event exhibits a very impulsive time profile coinciding with the non-thermal X-rays detected by the \Fermi Gamma-ray Burst Monitor~\citep[GBM;][]{Meegan_2009} detectors. Not only is the time profile of this flare very different from the other \Fermi LAT-detected BTL flares but also no coronal wave was observed in association with this event, setting it apart from 
the findings reported in \cite{Pesce-Rollins_2022}. 

In this work we present the observations of the solar flare of September 29, 2022, by \Fermi LAT and GBM together with those from the Spectrometer Telescope for Imaging X-rays (STIX), the Extreme-Ultraviolet Imager (EUI) Full Sun Imager (FSI) on board Solar Orbiter, observations from ground-based and space-borne radio spectrographs and the Nan\c{c}ay Radioheliograph, and EUV images from the Atmospheric Imaging Assembly on board the Solar Dynamics Observatory (SDO/AIA). In Section~\ref{sec:obs}, we review the observations and data analysis of this event. In Section~\ref{sec:otherbtl}, we compare the gamma-ray characteristics of this event with four other BTL events detected by the LAT as well as the impulsive phase of the on-disk event of September 6, 2011. In Section~\ref{sec:discussion}, we discuss the interpretation of the data from this flare in terms of flare-related particle acceleration and their propagation in large-scale magnetic structures.

\section{Observations and data analysis}\label{sec:obs}

\subsection{Instrumentation}

The Spectrometer Telescope for Imaging X-rays~\citep[][]{2020A&A...642A..15K} is an X-ray imaging spectrometer on board the Sun observing satellite Solar Orbiter~\citep[][]{2020A&A...642A...1M}. It observes solar thermal and non-thermal X-ray emission from 4 to 150 keV and provides quantitative information on the timing, location, intensity, and spectra of accelerated electrons as well as of high-temperature thermal plasmas. The EUI is a coronal imager on board Solar Orbiter and consists of three telescopes, the Full Sun Imager~\citep[EUI/FSI][]{SO_EUIpaper} and two high-resolution imagers that are optimized to image in Lyman-$\alpha$ and EUV (17.4 nm, 30.4 nm). For this flare, only FSI images at a 10-minute cadence are available. We note that all the times related to Solar Orbiter observations mentioned in this paper have been adjusted for the light travel time to Earth. 
 
From Earth's perspective, HXR and gamma-ray observations of this flare were made by the \Fermi GBM and LAT instruments. The GBM consists of twelve sodium iodide crystals for the 8 keV to 1 MeV range and two bismuth germanate crystals with sensitivity from 150 keV to 30 MeV. The LAT is an imaging gamma-ray detector designed to detect photons with energy from about 20 MeV  to over  300 GeV. The \Fermi\ satellite is in a low Earth orbit and is an astrophysics observatory, and therefore, it does not always have the Sun in its field of view.

Spectrographic observations at decimetric and longer radio wavelengths were provided by the ORFEES spectrograph in Nan\c{c}ay \citep{Ham:al-21} in the 144--1004 MHz frequency range, the e-Callisto spectrograph \citep{Ben:al-09} in Ooty (India, 50-170 MHz), 
the Nan\c{c}ay Decameter Array \citep[20-80 MHz; ][]{Lec:al-00}, and the WAVES spectrograph \citep{Bou:al-95} on the Wind spacecraft at decametric and longer wavelengths (14 MHz to about 3 kHz). Images at nine frequencies in the 151--432 MHz range were observed by the Nan\c{c}ay Radioheliograph (NRH) \citep{Ker:Del-97}, which consists of a T-shaped array (1.6 km $\times$ 1.25 km) with 48 antennas. Images in total intensity were taken with a 0.25-second cadence. 

\subsection{Observational overview}\label{sec:overview}
\begin{figure*}[ht!]
   \sidecaption
\includegraphics[width=0.9\textwidth,trim=0 60 0 0,clip]{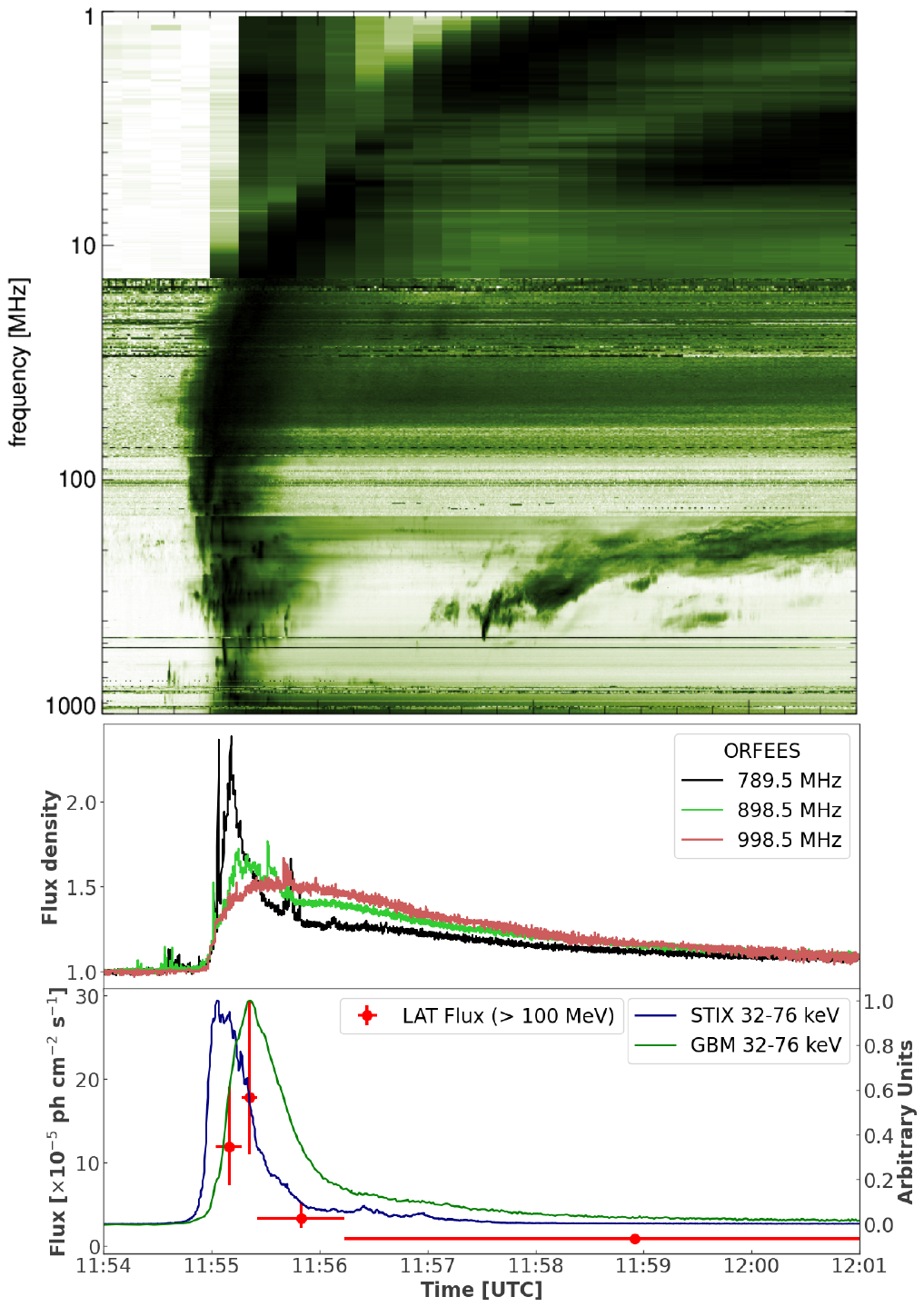}
\caption{Multiwavelength light curves of the flare of September 29, 2022. Top panel: Combined dynamic radio spectrum from ORFEES (144-1\,004~MHz), e-CALLISTO (81-144~MHz), NDA (13.9-80~MHz), and WAVES/RAD2 (1.1-13.8~MHz) showing a complex group of fast-drifting bursts with clear indications of repeated episodes of electron acceleration during the peak of the STIX and \Fermi LAT and GBM emission. Middle panel: ORFEES normalized flux density at three frequencies (789.5, 898.5, and 998.5 MHz). Bottom panel: \Fermi LAT $>$100 MeV flux points (red markers), normalized STIX and GBM time profiles in the 32-76 keV energy range are shown in blue and green solid lines, respectively. The Sun left the \Fermi LAT field of view at 12:01 UT.}
\label{fig:flux_lc} 
\end{figure*}

In Figure~\ref{fig:flux_lc} we report the multiwavelength light curves and dynamic radio spectra of the BTL flare of September 29, 2022. At 11:50 UT a C5.8 GOES class flare was detected off the eastern limb of the Sun with emission peaking at 12:01 UT and ending at 13:09 UT. The SoHO/LASCO CME catalog\footnote{\url{https://cdaw.gsfc.nasa.gov/CME_list/}} \citep{Yas:al-04} reports a broad CME with a velocity of 416 km/s. The Solar Orbiter-Earth separation was 178$^{\circ}$ at the time of the flare, with Solar Orbiter observing the east limb. Based on STIX imaging, the flare location as seen from Solar Orbiter is 16$^{\circ}$ behind the eastern limb as seen from Earth.

The ORFEES and e-Callisto radio spectrographs show a complex group of fast-drifting bursts from 11:55 to 11:56 UT, (type III, J; see below) emitted by electron beams in the corona. Their counterpart at frequencies below 100 MHz is a bright type III burst, which shows electrons traveling through the high corona. These emissions were followed by a type II radio burst visible from 11:57:30 and continuing past 12:15 UT essentially at frequencies above 100 MHz. Even though this type II burst is prominent, no EUV coronal wave was observed to be coincident with this flare. This is fairly uncommon, as type II bursts are typically associated with an EUV wave \cite[e.g.,][]{Kla:al-00}.

Both detectors on board the \Fermi\ satellite detected emission from the Sun starting at 11:55 UT. The LAT $>$100 MeV flux peaked at a value of 3.3$\pm$1.6$\times10^{-4}$ ph cm$^{-2}$ s$^{-1}$ between 11:55:16 -- 11:55:25 UT and remained significant until 12:01 UT, when the Sun left the field of view. The GBM 32-76 keV time profile reached its peak at the same time as the LAT flux, as can be seen in Figure~\ref{fig:flux_lc}. The LAT peak flux value for this event is comparable to the BTL flare of October 11, 2013 (SOL2013-10-11), and September 17, 2021 (SOL2021-09-17), both of which were located roughly 20$^{\circ}$ behind the eastern visible limb as well. No significant emission was observed from the Sun when it came back into the field of view of LAT at 13:15 UT.

\begin{figure*}[ht!]   
\begin{center}
\includegraphics[width=\textwidth,trim=0 0 0 0,clip]{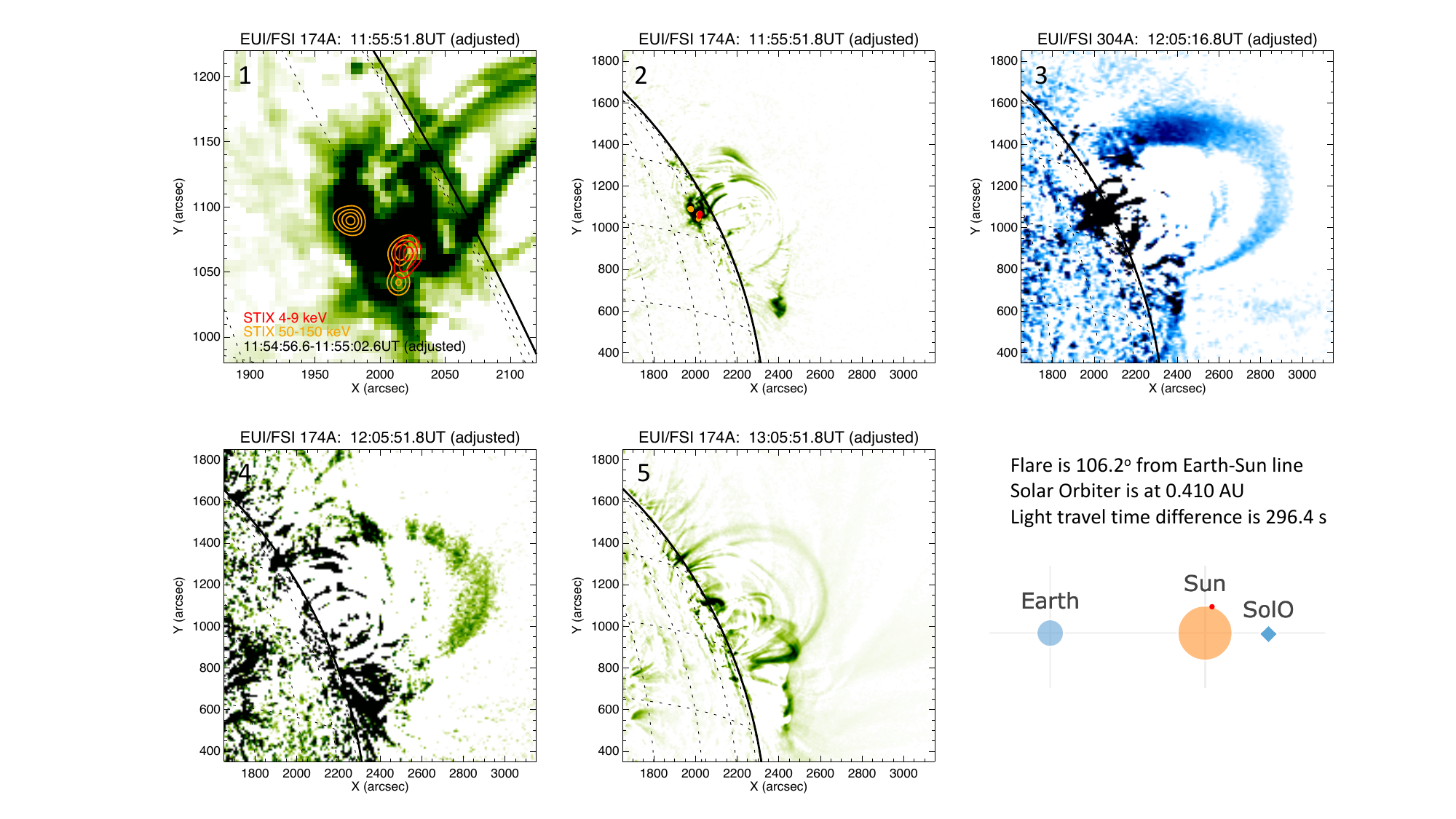}
\caption{STIX and EUI/FSI imaging observations of SOL20220929. Top row, from left to right: STIX thermal (red) and nonthermal (orange) sources overlaid on the 174 \angstrom\ EUI/FSI image at 11:55:51 UT  (panel 1); zoom out of  panel 1; and EUI/FSI 304 \angstrom\ image at 12:05:18 UT illustrating the presence of a very large loop (panel 3).  Bottom row, from left to right: EUI/FSI 184 \angstrom\ at 12:05:51 UT (panel 4); EUI/FSI 174 \angstrom\ image at 13:05:51 illustrating how the large loop is still present (panel 5).  The rightmost insert gives the position of Solar Orbiter and Earth relative to the flare site (indicated by the red dot). }
\label{fig:SOimaging} 
\end{center}   
\end{figure*}

\subsection{STIX data analysis}\label{sec:stix_analysis}

The Spectrometer Telescope for Imaging X-rays was in nominal science mode during the period of this flare. The dynamical binning in time quickly reached the highest cadence of 0.5 seconds, and essentially all of the impulsive phase of the flare was recorded at the highest cadence. Based on the time evolution of the emission measure and temperature derived from the STIX data, the estimated GOES class is $\approx$X2, with a lower limit of $\approx$X1 and a higher limit of $\approx$X4. After pre-flare background subtraction, the observed GOES class from the Earth viewing perspective is $\approx$C5. Hence, GOES only observed a few percent of the total soft X-ray flux due to the high occultation of this flare as seen from Earth. As the flare was at X class level and Solar Orbiter was rather close to the Sun at 0.41AU, the STIX attenuator was inserted during the impulsive phase of the flare. However, this had only a minor effect on the time profile, as shown in Figure~\ref{fig:flux_lc}. To correct for detector livetime, we used the best available livetime parameters as of February 2023.

The imaging of the flare from STIX and the EUI/FSI and Solar Orbiter's position relative to the Sun are shown in Figure~\ref{fig:SOimaging}. We used the CLEAN algorithm~\citep{1974A&AS...15..417H} and the standard software in SSWIDL (version v0.3.1) to produce the STIX images. The thermal source of this flare (red contours) is rather compact, outlining a flare loop with non-thermal sources at its footpoints (orange contours). These sources correspond to the standard flare picture where electrons precipitate to the footpoints producing non-thermal bremsstrahlung emission followed by ablation of heated chromosphere that fills the flare loop. However, for this flare, there is clearly a third non-thermal source visible at a strength similar to the non-thermal sources at the footpoints of the flare loop (as can be seen in the upper-left hand panel of Figure~\ref{fig:SOimaging}). The third source is clearly separated from the flare loop to the northeast, and it is not associated with a thermal X-ray source. This type of flare is generally believed to be associated with interchange reconnection\citep[e.g.][]{Krucker_2011}, where emerging magnetic flux reconnects with open field lines. The reconnection process changes the connectivity of one of the footpoints to an open field line (or an apparently open field line that connects back to the Sun far away from the flare). The third source is associated with this newly open field line. The lack of thermal X-ray emission is because heated plasma expands and escapes along the open field line, making the thermal X-ray signal much weaker compared to the flare loop that confines the heated plasma within its rather compact volume. The available EUI/FSI images are saturated at the flare site, but they show enhanced emission from field lines connecting the flare site with the high corona. 

\begin{figure*}[htbp] 
   \centering
   \includegraphics[width=0.45\textwidth]{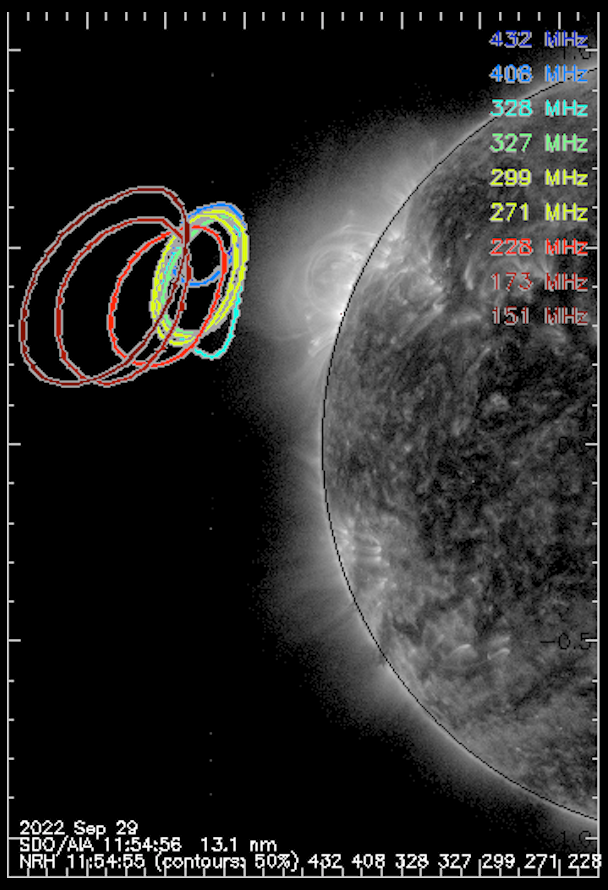} 
  \includegraphics[width=0.45\textwidth]{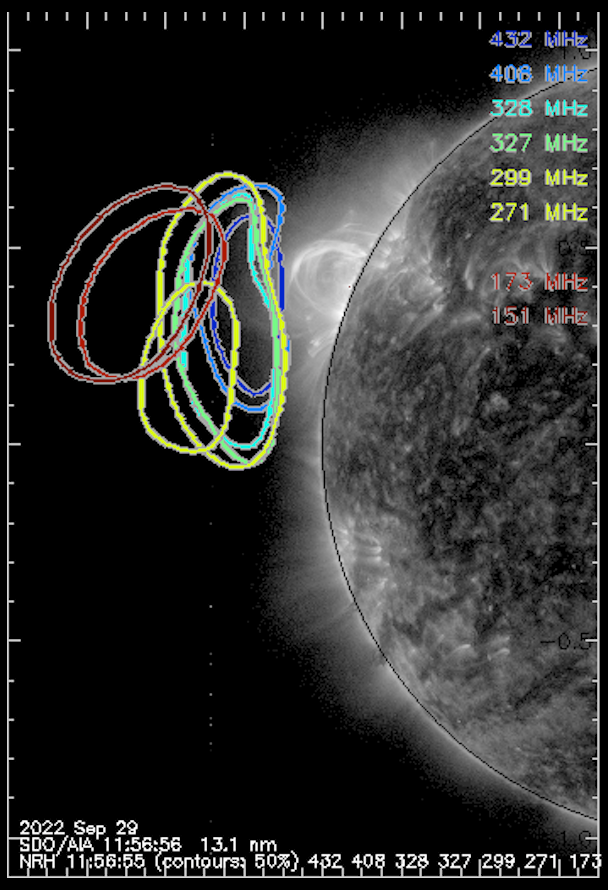} 
   \caption{Snapshots of SDO/AIA EUV images at 131 \AA\ near the start and the end of the impulsive phase radio emission with superposed isointensity contours of the radio sources (NRH; 10 s integration). A movie showing the evolution over the time range from 11:54 to 12:04 UT accompanies this figure. The movie extends until 12:04:45 UT, that is, far beyond the impulsive phase, which is the focus of this paper.}
   \label{Fig_AIANRH}
\end{figure*}

\begin{figure*}[htbp] 
   \centering
  \sidecaption
   \includegraphics[width=0.75\textwidth]{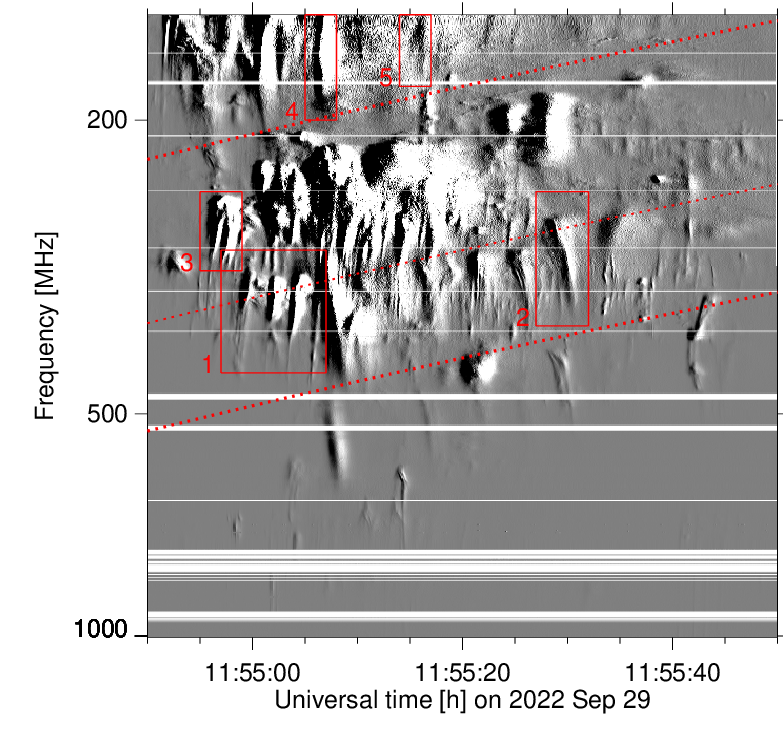} 
   \caption{Time-differenced ORFEES dynamic spectrum during the impulsive phase with three superposed dividing lines between chains of drifting bursts. The five numbered red rectangles delimit burst groups that were later used for the source localization. Time resolution is 0.1s.}
   \label{Fig_ORF20220929_DL}
\end{figure*}

\subsection{SDO/AIA and SUVI observations}
\label{sec:aia_suvi}

The EUV imagers in Earth orbit observed extended and strongly varying loops above the northeastern solar limb. We inspected multiwavelength EUV imagery of the corona with the Atmospheric Imaging Assembly \cite[AIA;][]{Lem:al-12} on board the Solar Dynamics Observatory \cite[SDO;][]{Pesnell:2012} and the GOES-16 Solar Ultraviolet Imager \cite[SUVI;][]{Darnel:2022}. These multi-band observations of the solar corona cover a broad range of temperatures, and they also profit from the high cadence of AIA (12 seconds) and the large field of view of SUVI (up to 2.3 R$_\odot$). The activation of this loop system above the northeast limb started around 11:55 UT and is very distinct in the AIA and SUVI filters that are sensitive to the high flare temperatures (94 and 131 \AA). The AIA and SUVI filters revealed the presence of an unusual high-flare loop system reaching up to about 120 Megameters above the solar limb (see, e.g., the frame at 11:56:55 UT, {\rd right} panel of Figure \ref{Fig_AIANRH}). This was  followed by an ejection of the upper part of the loop system, which then developed into the CME  (see also the movie accompanying Figure \ref{Fig_AIANRH}). There is one fast and localized ejection at the northern leg of the loops observed at 11:56--11:58 UT in AIA and SUVI, and a slower expansion of the higher lying flare loop system that started its rise at about 12:00 UT and left the AIA field of view at around 12:08 UT. The two snapshots in Figure\ \ref{Fig_AIANRH} show the AIA 131 {\AA} images at the beginning and end of the impulsive phase emission with superposed radio contours, which we discuss below.

In addition, a minor and very localized brightening was visible in the AIA 304 {\AA} and  1700 {\AA} 
filtergrams between 11:58 and 12:07 UT at N $23^\circ$E $60^\circ$. 
The observations suggest that there was an energy transport between the main activity behind the solar limb and certain regions on the disk, as seen from the Earth. 

\subsection{Radio spectral imaging as a tracer of electron acceleration and source geometry}

In this section we investigate in more detail the radio emission that accompanies the HXR and gamma-ray bursts. As seen in Figure \ref{fig:flux_lc}, the type III bursts below 100 MHz are accompanied by more complex spectral features in the 100--600 MHz range. At frequencies above 650 MHz (e.g., in the 790--1000 MHz range in the central panel), a broadband continuum dominates the radio spectrum. It extended throughout the microwave frequency range, as observed at fixed frequencies between 1.4 and 15.4 GHz by the San Vito station of the US Air Force Radio Solar Telescope Network (RSTN; data not shown), with similarly slowly-evolving time profiles as the 1000 MHz time history in Figure \ref{fig:flux_lc}. Bursts, which occur mostly at frequencies below about 650 MHz, are more clearly shown in the time-differenced ORFEES dynamic spectrogram displayed in Figure \ref{Fig_ORF20220929_DL}. Strong bursts  between about 11:54:50 UT and  11:55:30 UT accompanied the rise of the microwave emission and the rise, maximum, and early decay phase of the HXR burst observed by GBM. This suggests that the radio bursts are closely related to the acceleration time history of non-thermal electrons in general. Since the type II burst occurred well after the HXR and gamma-ray peaks, we do not go into a detailed study of it in this present paper. We note, however, that it is uncommon that the type II burst as well as the eruption (as observed by AIA; see Section~\ref{sec:aia_suvi}) occur after the impulsive flare phase.

The bursts in the ORFEES range have a frequency drift similar to type III bursts in absolute value (Figure \ref{Fig_ORF20220929_DL}). We therefore ascribed them to plasma emission by beams of non-thermal electrons. But while the bursts below 200 MHz are the high-frequency parts of the type III bursts observed by the Ooty (India) e-Callisto station, NDA, and Wind/WAVES, those bursts above about 200 MHz all have a sharp cutoff at low frequencies, and they display drifts both to lower and higher frequencies. The low-frequency cutoff means that the non-thermal electrons emit in a plasma with ambient electron densities above some threshold (i.e., in closed magnetic field structures). The negative and positive frequency drifts imply electron beams traveling toward lower and higher densities in different coronal structures. Overall, the bursts cluster in four slowly-drifting chains, which are clearly separated in the time-differenced dynamic spectrogram by three dividing lines. These dividing lines are overplotted on the dynamic spectrum as red dotted lines, and we refer to them as the low-frequency, central, and high-frequency dividing lines. They have been adjusted by eye to guide the discussion of the burst chains.

The central dividing line separates bursts of opposite drift, namely, bursts drifting toward high frequencies above the line and bursts drifting toward low frequencies below the line. This central dividing line is naturally interpreted as the signature of a region from where electron beams are released upwards (negative drift, i.e., toward lower frequencies) and downwards \cite[positive drift, i.e.,  toward higher frequencies; ][]{Asc-02,TBL:al-16,Kle:al-22}. The negatively drifting bursts on the low-frequency side of the dividing line are type J bursts. The drift toward lower frequencies slows down and stops at some frequency (see \citeauthor{ScR-Rat-14} \citeyear{ScR-Rat-14}, and references therein, \citeauthor{ZhaJ:al_23} \citeyear{ZhaJ:al_23}). This is generally explained by electron beams being in a magnetic loop, as they stop radiating when they reach the top of the loop. The low-frequency border of the bursts is seen to proceed stepwise toward lower frequencies, but it can be constant for some seconds (e.g., 11:55:07-11:55:20 UT at 220 MHz). 

On the high-frequency side of this burst group, Figure \ref{Fig_ORF20220929_DL} shows a few positively drifting bursts at frequencies above the high-frequency dividing line. The  burst at 11:55:07 UT in the band 480-620 MHz has a counterpart in the band 315-470 MHz, and the burst  at 11:55:29 UT in the band  410-520 MHz has a counterpart in the band 270-370 MHz. In both cases, the high-frequency burst is at about 1.5 times the frequency of the low-frequency counterpart. Because of the spectral similarity and the constant frequency ratio, we consider that this high-frequency drifting chain is closely related to the positively drifting bursts adjacent to the dividing line and that it is not evidence of an independent accelerator. This high-frequency drifting chain could be explained by the simultaneous downward release of electrons into two different structures with a ratio of ambient densities of $1.5^2$. 

The radio bursts below the low-frequency dividing line (i.e., below 200 MHz) comprise the high-frequency part of the type III bursts seen below 100 MHz. However, several of the bursts have high-frequency counterparts that drift in the opposite direction, similar to the situation around the central dividing line. This can be clearly seen for the burst in rectangle 4 in Figure \ref{Fig_ORF20220929_DL}. The separation between these oppositely drifting bursts below the low-frequency dividing line hence shows a different source of electron beam generation, namely, higher in the corona but closely related in time to the lower acceleration region where the X-ray emitting electrons are accelerated.

\begin{figure*}[htbp] 
   \centering
      \includegraphics[width=.45 \textwidth]{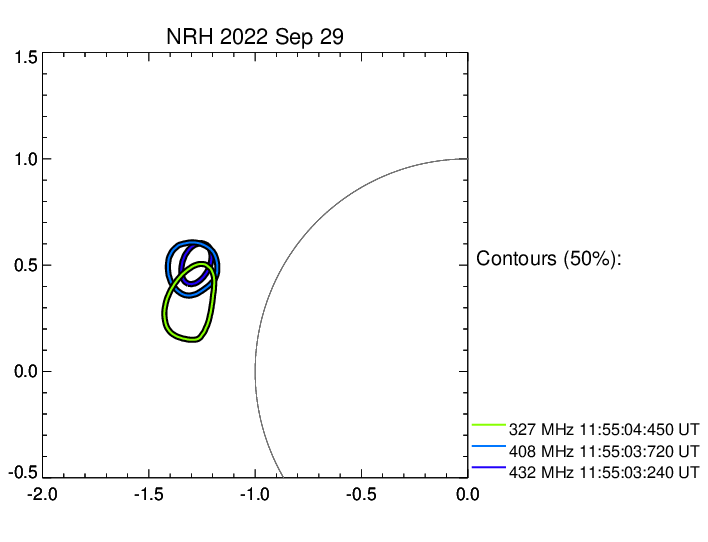} 
      \includegraphics[width=.45 \textwidth]{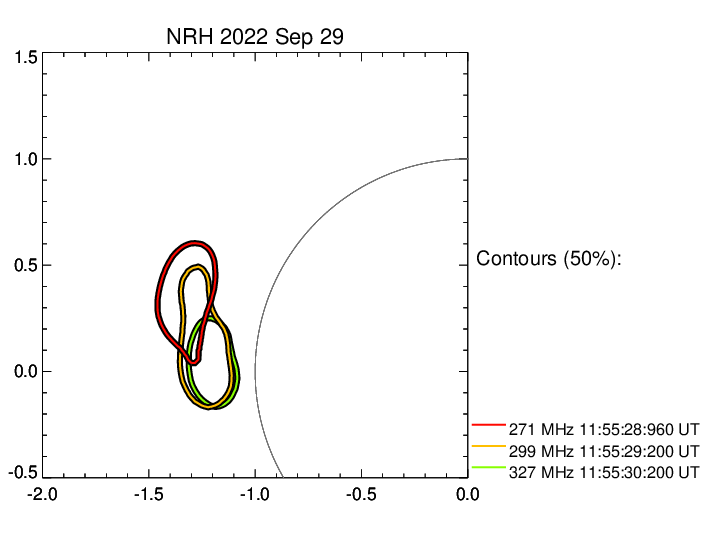} 
      \includegraphics[width=.45 \textwidth]{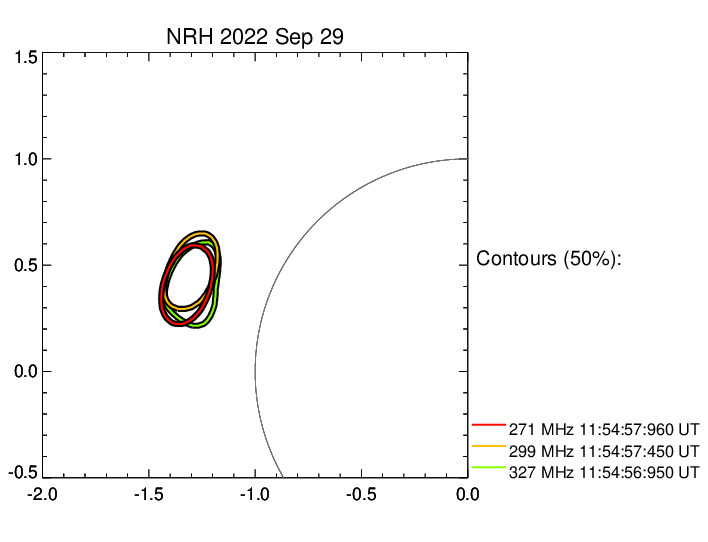} 
      \includegraphics[width=.45 \textwidth]{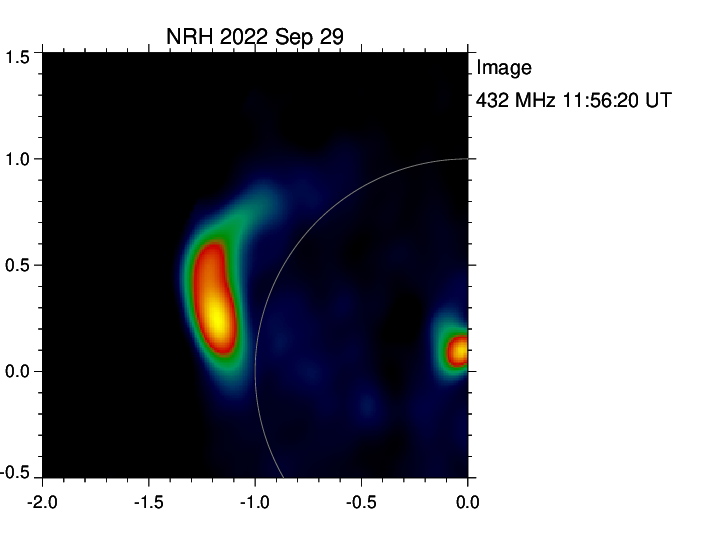}

   \caption{NRH maps during selected periods of the event. Top left: Isointensity contours at half maximum during the type J burst at 11:55:03 UT (rectangle 1 in Figure \ref{Fig_ORF20220929_DL}). Top right: Reverse-drifting burst near 11:55:30 UT (rectangle 2). Bottom left: Type J burst at 11:54:57 UT on the low-frequency side of the red spectral dividing line (rectangle 3). Bottom right: Image of the continuum source at 432 MHz (average 11:56:20 to 11:56:50 UT). 
}
   \label{Fig_NRH20220929}
\end{figure*}

We analyzed NRH images during individual bursts using the full time resolution of 0.25 s. The geometries of selected bursts are shown in Figure \ref{Fig_NRH20220929}, where sources at different frequencies are represented by isointensity contours at half maximum with different colors. The upper-left panel shows sources of the early type J burst on the high-frequency side of the spectral dividing line (rectangle 1 in Figure \ref{Fig_ORF20220929_DL}). The source at 327 MHz (green contour) is located in projection to the south of the 408 and 432 MHz sources (blue contours). Since the emission is at the plasma frequency or its harmonic, the height of the sources is expected to increase with decreasing frequency. The southward location of the source at the lower frequency is therefore a projection effect, which can be understood if the parent magnetic structure connects the active region behind the solar limb, where the electron beams are supposed to originate, with a point located southwestward of the active region on the Earthward solar hemisphere. The 327 MHz source, which the dynamic spectrum shows to be close to the loop top, is also more elongated than the one at 408 or 432 MHz. The elongation can be explained by the small density gradient along the magnetic field line around the top of the magnetic structure. 

The type J bursts are followed by a chain of mostly positively drifting bursts on the high-frequency side of the central dividing line of Figure \ref{Fig_ORF20220929_DL}. The sources of the burst with the best NRH frequency coverage near 11:55:30 UT (rectangle 2 in Figure \ref{Fig_ORF20220929_DL}) are displayed in the top-right panel of Figure \ref{Fig_NRH20220929}. The high-frequency source (327 MHz) is now southwestward of the low-frequency sources (271 and 299 MHz). These sources are located in the other leg of the magnetic structure that hosted the previous type J burst. This geometry is again consistent with the magnetic field lines from the parent active region behind the limb to the disk. The positively drifting burst is observed in the Earthward leg of this structure.

The sources of a well-observed type J burst on the low-frequency side of the spectral dividing line (rectangle 3 in Figure \ref{Fig_ORF20220929_DL}) are displayed in the bottom-left panel of Figure \ref{Fig_NRH20220929}. They are again on the northern side of the magnetic structure identified at higher frequencies (top row of the figure). The sources at different frequencies appear superposed on each other, which means that the magnetic field lines that guide the electron beams are parallel to the line of sight, consistent with the idea that the structure guiding these electron beams overlies the one hosting the bursts on the high-frequency side of the central dividing line. 

The bottom-right panel of Figure \ref{Fig_NRH20220929} is a color-scale image of the radio continuum source at 432 MHz after the end of the bursty emission. The image was integrated between 11:56:20 and 11:56:50 UT, after filtering out fluctuations that come from residual weak bursts and from sidelobes of the noise storm source seen near the disk center. The continuum radio source was observed at the time of the last significant excess measured by LAT (Figure \ref{fig:flux_lc}). 

In summary, the imaging observations around the central dividing line in the dynamic spectrum show emission from an extended magnetic structure bridging the solar limb. The northern leg of the radio structure projects to above the northern leg of the loop in the AIA image (Figure \ref{Fig_AIANRH}), where a fast localized ejection was observed in EUV images (Section \ref{sec:aia_suvi}).  The source geometry during individual bursts turns out to be well represented by the radio contours in Figure \ref{Fig_AIANRH}, where a much coarser resolution of 10 s was used. This figure also shows the sources at lower frequencies (228, 173, 151 MHz) than discussed above, which lie above the northern part of the high-frequency sources, again connected to the region where the strongest activity is observed in the EUV images. The sources at 151 and 173 MHz are the sources of the type III bursts near their start. These sources are hence close to the region where electron beams are released onto field lines that are open to the interplanetary space. The oppositely drifting bursts that appear on their high-frequency side are not imaged by the NRH. A possible scenario explaining the observed source geometry and spectral organization is discussed in Sect. \ref{sec:discussion}. 


\subsection{Connection between behind-the-limb and on-disk HXR emission}
\label{sec:comparison_hxr}

\begin{figure}[ht!]   
\begin{center}
\includegraphics[width=0.45\textwidth,trim=0 0 0 0,clip]{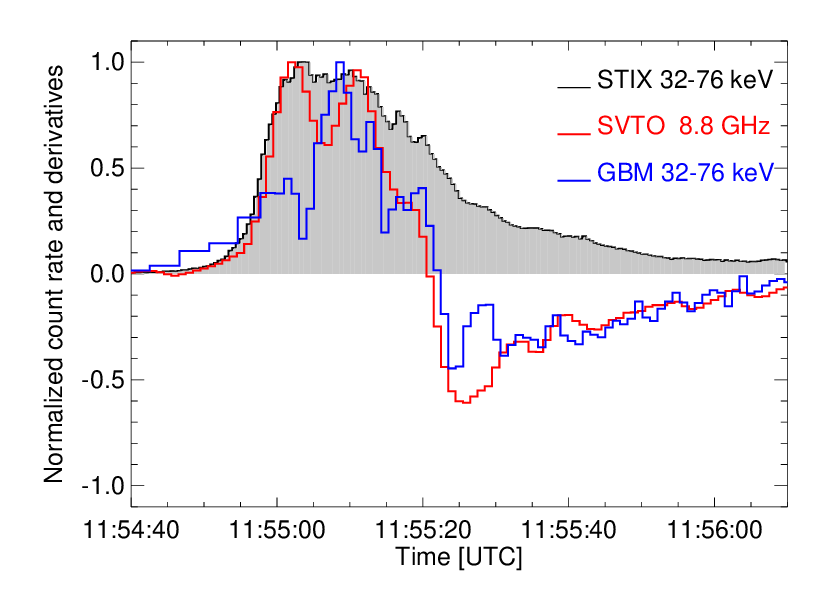}
\caption{Time histories of the HXR count rate observed by STIX (black curve and gray-shaded background) of the derivatives of the count rate observed by GBM (blue) and of the microwave flux density at 8.8 GHz (red; San Vito station of RSTN). Each curve was normalized to its maximum.}
\label{Fig_HXRcomp} 
\end{center}   
\end{figure}

The time history of the HXR emission observed by STIX shows a series of bursts, while GBM observed a very smooth time profile (bottom panel of Figure \ref{fig:flux_lc}), as is usual for occulted X-ray sources \citep[e.g.,][]{Kan:al-92}. This is because the bremsstrahlung process leads to bright emission in the chromospheric thick-target sources seen by STIX, where the non-thermal electrons lose their energy by collisions so rapidly that particle injections become discernible as individual bursts. 

The smooth time profile seen by GBM can be understood if the details of the electron releases are smeared out by trapping and thin-target emission. In order to emphasize the changes in the GBM time profile, in Figure \ref{Fig_HXRcomp}, we compared its derivative (blue line) with the STIX count rate time history (black). The derivative of the microwave flux density at 8.8 GHz observed by the RSTN San Vito station is also shown in the figure. It appears that the three major distinct components of the STIX time profile (start to 11:55:03 UT, 11:55:03 to 11:55:15 UT, and 11:55:15 UT to the end) have counterparts in the derivatives of the count rate and flux density time histories seen from the terrestrial viewpoint. These similarities suggest that the electrons seen by STIX and from the terrestrial viewing direction come from the same processes of acceleration.

\subsection{\Fermi GBM and LAT data analyses}{\label{sec:lat_analysis}}
The GBM data was background subtracted using a polynomial fit to the background before and after the flare to approximate the time variation during the flare. After background subtraction and normalization of the count rates, the time profiles of the five most sunward directed detectors were compared. As all of these profiles agree well with each other, we used the summed profiles of the five most sunward directed detectors. 

We performed an unbinned likelihood analysis of the LAT data within the Multi-Mission Maximum Likelihood (\texttt{3ML})\footnote{\url{https://threeml.readthedocs.io/en/stable/index.html}} framework using \texttt{fermitools}\footnote{\url{https://github.com/fermi-lat/Fermitools-conda/wiki}} version 2.0.8. We selected P8R3\_SOURCE\_V3 class events from a 10$^{\circ}$ circular region centered on the Sun and within 100$^{\circ}$ from the local zenith (to reduce contamination from the Earth limb). We first analyzed the flare by integrating over the entire time interval that the Sun was in the field of view, from 11:55:02 -- 12:01:36 UT, and selecting all the events above 60 MeV to better constrain the shape of the spectrum at low energies and test three models to the \Fermi LAT data. The first two, a pure power law (PL) and a PL with an exponential cutoff (PLEXP),\footnote{The definition of the models used can be found here: \url{https://fermi.gsfc.nasa.gov/ssc/data/analysis/scitools/source_models.html}.} are phenomenological functions that may describe bremsstrahlung emission from relativistic electrons. The third model uses templates based on a detailed study of the gamma rays produced from the decay of pions originating from accelerated protons with an isotropic pitch angle distribution in a thick-target  model \citep[updated from][]{murp87}. In all three analyses, the background was modeled by a fixed contribution coming from the galactic gamma-ray emission (described by the standard template available in \texttt{fermitools}) and by an isotropic emission describing the unresolved particle background (also described by the standard available template). This latter background component was left free to vary, as it encompasses the background variation due to orbital modulation.

We relied on the likelihood ratio test and the associated test statistic \citep[TS;][]{Mattox:96} to estimate the significance of the detection. The TS of the PL fit (TS$_{\rm PL}$) indicates the significance of the source detection under the assumption of a PL spectral shape, and the $\Delta$TS=TS$_{\rm ALT}$ - TS$_{\rm PL}$ quantifies how much an alternative model improves the fit. We note that the significance in $\sigma$ can be roughly approximated as $\sqrt{\rm TS}$. For the PL model, we obtained TS$_{\rm PL}$=238, while for the exponential cutoff, we obtained an improvement of $\Delta$TS$\approx$4, suggesting that the curved model is preferred with $\sim$2$\sigma$ significance. 

In order to help define the optimal binning for a time-resolved analysis, we used the results from the time-integrated likelihood analysis using the exponential cutoff model to compute the probability that each event is associated with the Sun (using the \texttt{gtsrcprob} tool available in \texttt{fermitools}). We then required that each bin contain at least 10 photons and that the start of the detection coincide with the arrival of the first photon, with a probability greater than 0.9 of being associated with the Sun. The selected bins together with the results for the time-resolved likelihood analysis are reported in Table~\ref{tab:spectral_results}. Since the PLEXP model is only marginally preferred, we performed the time-resolved analysis with the PL model.

\begin{table}[h!]
   \caption{Results for the time-resolved likelihood analysis.}
    \centering
    \begin{tabular}{c r  r r}
    Time window & Flux$_{0.1-1.0\rm GeV}$ & Photon index & TS$_{\rm sun}$ \\
          &  cm$^{-2}$ s$^{-1}$   & &   \\
    \hline
     11:55:02 -- 11:55:16 & $ 12 \pm 6 $  & $ -4.8 \pm 0.9 $  & 77\\
    11:55:16 -- 11:55:25 & $ 18 \pm 9 $  & $ -5.3 \pm 1.0 $  & 116\\
    11:55:25 -- 11:56:13 & $ 3.3 \pm 1.6 $  & $ -5.3 \pm 1.0 $ &  70\\
    11:56:13 -- 12:01:36 & $ 0.9 \pm 0.4 $  & $ -4.4 \pm 0.8 $  & 33\\

    \hline
    \end{tabular}
    \tablefoot{The analysis was performed using the PL model ($\frac{dN(E)}{dE}=N_0E^{\Gamma}$). In each interval, we report the time window for the analysis, the photon flux between 100~MeV and 1~GeV (in $10^{-5}$cm$^{-2}$ s$^{-1}$), the photon index, and the significance of the source in TS. }
    \label{tab:spectral_results}
\end{table}

\section{Comparison with other \Fermi LAT behind-the-limb and on-disk flares}
\label{sec:otherbtl}
One of the most striking aspects of this flare is how the $>$100 MeV flux peaks in coincidence with the non-thermal GBM flux and how both of these emissions peak within less than 20 seconds of the STIX emission (see Figure~\ref{fig:flux_lc}). The shared peak time between LAT and GBM has never been observed in a BTL flare before and is suggestive of a common origin. Other characteristics of this flare that set it apart from the previously detected BTL flares are the fast rise and decay time. In order to facilitate the comparison between this flare and the other BTL detected by the LAT, we estimated the rise time of the gamma-ray time profile and the time difference between the peak flux in gamma rays and non-thermal GBM emission ($\Delta$Peak hereafter). The rise time was taken to be the difference between the start of the first detection to the midpoint of the time bin in which the flux peaks, whereas $\Delta$Peak is simply the time difference between the peak flux in 32-76 keV and $>$100 MeV as observed by GBM and LAT, respectively.

\begin{figure}[ht!]   
\begin{center}
\includegraphics[width=0.5\textwidth,trim=2cm 0 0 0,clip]{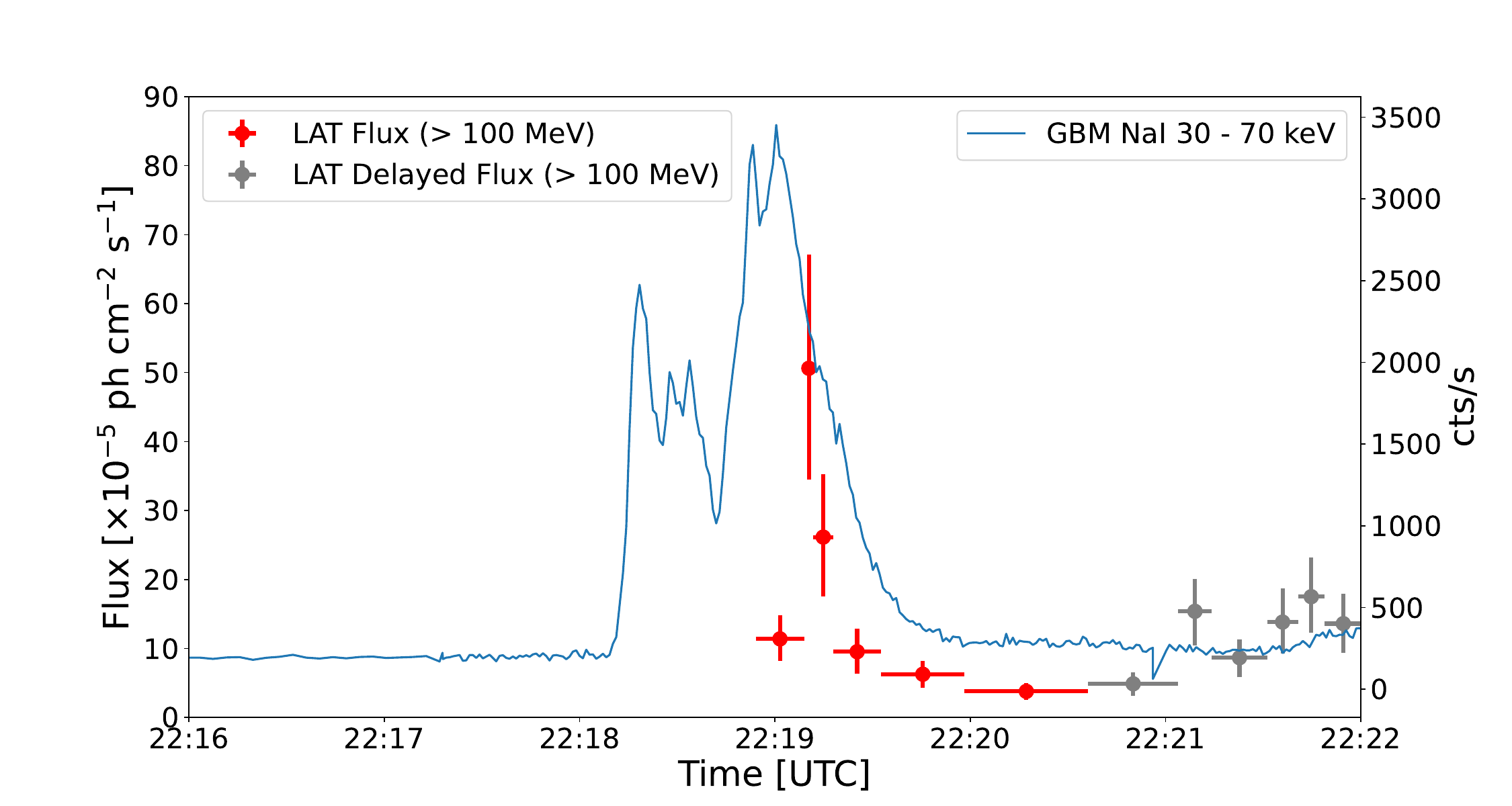}
\caption{Multiwavelength light curves of the on-disk flare SOL2011-09-06. The \Fermi LAT $>$100 MeV flux points are indicated with red markers, and the GBM time profiles in the 32-76 keV energy range are in blue. The SOL2011-09-06 also had a delayed emission component, indicated by the gray points following 22:22 UT. The Sun came into the field of view at 22:11 UT, and prior to 22:19 UT, no significant detection of the flare was observed by \Fermi LAT.}
\label{fig:flux_ondisk_lc} 
\end{center}   
\end{figure}

We found that the rise time for the BTL flares are all on the order of minutes, whereas the flare of September 29, 2022, is only 18$\pm$4 seconds, and the value of $\Delta$Peak is only 5$\pm$4 seconds, when the other BTL flares all peak several minutes after the GBM non-thermal peak. The peak flux value for this event lies within those of the other BTL flares. The fast rise time and small value of $\Delta$Peak are characteristics reminiscent of impulsive on-disk flares.  
We searched through the sample of impulsive on-disk flares listed in Table 3 of \cite{flarecatalog_2021} for events with sufficient emission above 100 MeV to be able to perform a time-resolved analysis in more than four time windows and with a single peak. While there are several flares with significant high-energy emission, the flare that best suited our requirements for this case study was the flare of September 6, 2011 (SOL2011-09-06). This flare, classified as an X2.1 GOES class, erupted from active region (AR) NOAA 11283 at 22:12 UT and was associated with a CME with an estimated speed of 990 km/s \citep{Dissauer:2016} based on STEREO-A COR data, and the CME was observed almost perfectly on the plane of sky. The position of the AR from Earth view at the time of the flare was N14W18. It was a bright impulsive flare observed and studied in several wavelengths~\citep{2013ApJ...765...37F,Xu_2014,SunQuakes}, and \cite{SunQuakes} reported that the HXR contours as measured by RHESSI show two closely spaced sources, as expected to occur for the footpoints of a flare loop.
We performed an unbinned likelihood analysis on SOL2011-09-06 following the same procedure described in Section~\ref{sec:lat_analysis} with the exception that for this flare, we used P8R3\_TRANSIENT020E\_V2 class events instead of P8R3\_SOURCE class (in order to increase the statistics). The LAT time-resolved light curve is shown together with the GBM 32-76 keV time profile in Figure~\ref{fig:flux_ondisk_lc}. The flare SOL2011-09-06 had a very bright $>$100 MeV delayed emission component that lasted until 22:47, when the Sun left the field of view of the LAT. For the purposes of this study, we only considered the impulsive phase of this event, and this is indicated by the red points in Figure~\ref{fig:flux_ondisk_lc}. We defined the end of the impulsive phase as the time where the flux value reaches its minimum following the initial peak.\footnote{We note that the exact definition of the end of the impulsive phase and the start of the delayed phase is not relevant for the comparison in this work.} In Table~\ref{tab:flare_characteristics}, we report the values of the rise time, $\Delta$Peak, and peak flux for five LAT-detected BTL flares together with the impulsive phase of SOL2011-09-06. We used the results reported by \cite{Pesce-Rollins_2022} and \cite{Ackermann_2017} to estimate the values for the BTL flares listed in the table.

From this table it is clear that the flare SOL2022-09-29 is remarkably similar in terms of the basic characteristics listed in the table to the on-disk impulsive phase of the flare SOL2011-09-06. Another striking similarity between the flare of September 29, 2022, and SOL2011-09-06 is that in less than a minute the peak flux dropped  by an  order of magnitude. The BTL flares of SOL2013-10-11, SOL2014-09-01, and SOL2021-07-17 all took more than 10 minutes for the peak flux to drop by the same amount, whereas for the BTL flare of SOL2021-09-17, the Sun left the LAT field of view 4 minutes after the peak, and the flux had still not decayed by an order of magnitude. This simple comparison suggests that the BTL event of September 29, 2022, and the on-disk flare SOL2011-09-06 are driven by an acceleration mechanism  or governed by transport processes that {could be} different from those at work in the remaining four BTL flares listed in Table~\ref{tab:flare_characteristics}.

\begin{table*}[h!]
  \centering
   \caption{Comparison between BTL and on-disk LAT flares.}
    \begin{tabular}{c| c c c c}  
    Flare & Rise time & $\Delta$Peak & Peak Flux & AR\\
             & (min:sec) &  (sec)   & (10$^{-5}$ ph cm$^{-2}$ s$^{-1}$) &\\
    \hline
    \hline
      SOL2013-10-11 & 9:00 $\pm$1:00   & 600 $\pm$ 60 & 49$\pm$2 & N21E103 \\
     SOL2014-09-01 & 9:00 $\pm$1:00  & 240 $\pm$ 60  & 565$\pm$21 & N14E126 \\
     SOL2021-07-17 & 8:00 $\pm$1:00  &  -  & 4$\pm$1 & S20E140 \\
     SOL2021-09-17 & 1:09 $\pm$0:07  & 189 $\pm$ 7 & 67$\pm$20 & S30E100\\
     SOL2022-09-29 & 0:18 $\pm$ 0:04 & 5 $\pm$ 4  & 44$\pm$14 & N26E106\\
        \hline
        \hline
    SOL2011-09-06 & 0:16 $\pm$ 0:01 & 10 $\pm$ 1 & 50$\pm$16 & N14W18
    \end{tabular}
      \tablefoot{Rise time, difference between GBM 32-76 keV and $>$100 MeV peak times ($\Delta$Peak), peak flux value {\rd of the gamma-ray emission}, and the position of the active region for the five BTL flares detected with the LAT and with sufficient statistics to perform a time-resolved analysis (top table) and the impulsive on-disk flare of September 6, 2011 (bottom table). The criteria for the selection of the on-disk flare is described in the text. The BTL flare that is not considered here is the flare of January 6, 2014, due to a lack of sufficient coverage by the LAT. We do not list the value of $\Delta$Peak for SOL2021-07-17 because the GBM did not detect any significant emission from this event.}
    \label{tab:flare_characteristics}
\end{table*}

\section{Discussion}
\label{sec:discussion}
In this work we reported on the multiwavelegth observations of the BTL solar flare of September 29, 2022. The greater than 100 MeV time profile of this event had a fast rise and decay that is reminiscent of impulsive on-disk gamma-ray flares. The LAT and GBM emission peaked at the same time (within the statistical uncertainties) and within 20 seconds after the STIX 32-76 keV emission peaked. 
We compared the basic characteristics of the gamma-ray time profile (rise/decay times and the time delay between the gamma-ray peak flux and the peak flux in HXRs) of this event with those of four BTL flares observed by LAT and with an on-disk event, and we found that the flare of September 29, 2022, shares more similarities with the on-disk flare than with the other BTL LAT-detected flares. \cite{1999A&A...342..575V}  came to a similar conclusion for an event producing electron bremsstrahlung emission up to several tens of MeV, but this is the first case in the \Fermi LAT energy range. In addition, complementary data allowed us to establish a scenario of the relationship between the flare behind the limb and gamma-ray emission seen from Earth's orbit.

The radio observations show several bursts of type J and reverse drift above 200 MHz, type III bursts below 200 MHz during the impulsive phase of the gamma-ray flare, and a type II burst  
during the later phase of the event. The type II burst indicates that a shock is present during the event. However, its onset occurred during the decay phase of the $>$100 MeV emission. This shock wave cannot be the source of the LAT emission. Contrary to the findings reported in \cite{Pesce-Rollins_2022}, no coronal wave was observed to be associated with this event. As was described in Section~\ref{sec:aia_suvi}, the similarity between the STIX time histories and the derivatives of the GBM count rates suggests that the electrons producing the emission observed by STIX and by GBM were accelerated by the same process. Both EUI/FSI and SDO/AIA imaging show the presence of very large loops seen from the Solar Orbiter and Earth viewing perspectives.

However, the delay of approximately 20 seconds between the HXR peak seen by GBM and STIX cannot be explained via the simple picture that electrons escape the flare site and 
{\rd
freely stream
} 
to the visible side of the Sun because the distances required would be greater than four solar radii.\footnote{The electrons that produce the observed emissions are typically semirelativistic. If they travel at half the speed of light, a delay of $\approx$20s implies a traveled distance of over four solar radii.} The structures observed in radio for this event are much smaller than this. 
{\rd
Moreover,
} 
the smoothness of the HXR and microwave time profiles observed from Earth suggests that 
{\rd
the electrons do not stream freely but are trapped. This
} 
can result in a time delay. To estimate its size, we assumed that the STIX time profile is the injection function and that the HXR emission is related to the number of electrons injected and trapped in a magnetic structure\footnote{We assumed the electrons were perfectly trapped.} \cite[See for example][]{1982A&A...108..306V}. Therefore, any delay between the time integral of the injection function and the STIX count rate will be related to the induced delay due to trapping. We found that the delay between the half-maximum values is 17 s, that is, of the same order as the observed delay. 

The imaging observations of the radio sources during the HXR and gamma-ray emission suggest that electrons are injected into large-scale closed magnetic structures that bridge the northeastern solar limb, connecting the occulted flare with the Earthward hemisphere of the Sun. The dynamic spectrum comprises two groups of bursts above and below a spectral dividing line around 200 MHz, with a combination of forward-drifting and reverse-drifting bursts in each group. The existence of two groups indicates two regions of electron release at different altitudes. While the type III bursts in the low-frequency group come from electrons released to the high corona and interplanetary space, the sharp low-frequency cutoff of the high-frequency group (type J bursts) demonstrates the release into closed magnetic structures. At the same time, electrons were injected downward (reverse drift) into the low atmosphere, where they generated HXR emission. Such sharp radio cutoffs were reported in earlier gamma-ray events, too  \citep{Tro:al-98,Rgr:al-99}. In addition, a localized brightening was observed in the SDO/AIA 304\AA\ and 1700\AA\ channels, suggesting that energy was being transported between the main activity behind the visible limb and the disk as seen from Earth during the flaring activity.

In a standard scenario of a CME eruption \citep[see review by][]{Che-11}, a magnetic flux rope rises in the corona, forming a current sheet below where magnetic reconnection creates a loop arcade \citep[see][for detailed case studies of current sheets in events where high-energy particles are accelerated]{Asc:Alx-01,Wrr:al-18,Cai:al-19}. Another current sheet is expected at greater altitude, where the rising flux rope pushes against the arcades of overlying magnetic field lines. This scenario can explain the observed spectral organization of radio bursts during the event, namely, the existence of two groups of bursts with oppositely directed drifts. Bursts at  high frequencies with oppositely directed drifts on either side of the central spectral dividing line can be ascribed to electrons released from the lower current sheet upward into the flux rope and downward into the loop arcade, where they are trapped or precipitated into the dense lower-atmosphere emitting HXR. The bursts in the low-frequency group, below the low-frequency dividing line, would be ascribed to electrons that are either accelerated at the upper current sheet or released from the flux rope as it reconnects with the ambient magnetic field. If this field is open, the electrons will generate type III bursts. The overall drift of the dividing lines toward lower frequencies reflects the rise of the current sheets with the erupting flux rope. The stepwise decrease of the low-frequency cutoff of the high-frequency burst group suggests that this rise occurs in discrete episodes rather than continuously. All of these observational results support the scenario that the $>100$ MeV emission originates from particles accelerated in large loops or during the formation and evolution of a magnetic flux rope that connects the flaring site behind the limb to the visible disk of the Sun.

 {\rd Based on the observations of six BTL flares with measured energies greater than 100 MeV, we can conclude that there are more ways than one for the accelerated particles to travel from the acceleration site to the visible disk. 
 {\rd
The observations demonstrate that a gamma-ray flare may involve complex magnetic structures extending well beyond the parent active region. 
 }
In \cite{Pesce-Rollins_2022}, observational evidence was reported in favor of the CME-shock scenario for four out of the five BTL flares considered in that work. In contrast, the flare reported in this work clearly indicates that the CME shock is not the source of the gamma-ray emission. {In fact, the majority of the gamma-ray emission of this event occurred several minutes prior to the appearance of the type II burst, and no significant variation in the time profile was observed in coincidence with the onset of the type II burst. Therefore, this suggests that there is no indication of shock acceleration of relativistic ions. \cite{Man:al-22} argued recently that the acceleration of type II burst emitting electrons may be due to shocks with a relatively low Mach number, whereas relativistic particles require particularly high Mach numbers \citep[see the model by][which applies to relativistic protons]{Afa:al-18b}.}
 
Our comparisons between the LAT-detected BTL flares also suggest that the flares with a clear association with the CME shock have a significantly longer duration than the September 29, 2022 event.
While indications of trapping are present in the event studied here, it could not have lasted very long, judging from the duration of the HXR event observed from Earth's orbit, {setting this event apart from the remaining BTL flares observed with \Fermi LAT.}
  
This work emphasizes processes that do contribute to the broad variety of gamma-ray events, but we cannot pretend to give a general conclusion 
on the overall population of BTL flares or gamma-ray flares in general. {However, with the observation of the flare of September 29, 2022, we can say that the CME-driven shock is not a necessary condition for BTL gamma-ray flares.}}

\begin{acknowledgements}
The \textit{Fermi} LAT Collaboration acknowledges generous ongoing support
from a number of agencies and institutes that have supported both the
development and the operation of the LAT as well as scientific data analysis.
These include the National Aeronautics and Space Administration and the
Department of Energy in the United States, the Commissariat \`a l'Energie Atomique
and the Centre National de la Recherche Scientifique / Institut National de Physique
Nucl\'eaire et de Physique des Particules in France, the Agenzia Spaziale Italiana
and the Istituto Nazionale di Fisica Nucleare in Italy, the Ministry of Education,
Culture, Sports, Science and Technology (MEXT), High Energy Accelerator Research
Organization (KEK) and Japan Aerospace Exploration Agency (JAXA) in Japan, and
the K.~A.~Wallenberg Foundation, the Swedish Research Council and the
Swedish National Space Board in Sweden.
Additional support for science analysis during the operations phase is gratefully
acknowledged from the Istituto Nazionale di Astrofisica in Italy, the Centre
National d'\'Etudes Spatiales in France. This work performed in part under DOE
Contract DE-AC02-76SF00515.

Solar Orbiter is a space mission of international collaboration between ESA and NASA, operated by ESA. The STIX instrument is an international collaboration between Switzerland, Poland, France, Czech Republic, Germany, Austria, Ireland, and Italy. The SO/EUI instrument was built by CSL, IAS, MPS, MSSL/UCL, PMOD/WRC, ROB, and LCF/IO with funding from the Belgian Federal Science Policy Office (BELSPO/PRODEX PEA 4000134088); the Centre National d'Etudes Spatiales (CNES); the UK Space Agency (UKSA); the Bundesministerium für Wirtschaft und Energie (BMWi) through the Deutsches Zentrum für Luft- und Raumfahrt (DLR); and the Swiss Space Office (SSO).

Radio data were provided by the Nan\c{c}ay Radio Observatory via the Solar Database RSDB, by the e-Callisto consortium and the Wind/WAVES team. The LASCO CME catalog is generated and maintained at the CDAW Data Center by NASA and The Catholic University of America in cooperation with the Naval Research Laboratory. SOHO is a project of international cooperation between ESA and NASA. KLK acknowledges support by the French solar-terrestrial physics programme PNST of CNRS and the polar institute IPEV. SK is supported by the Swiss National Science Foundation Grant 200021L\_189180 for STIX.

MPR would like to thank Ed Cliver for the insightful comments and suggestions provided.
\end{acknowledgements}

%

\bibliographystyle{aa}
\bibliography{SOL220929,klk}{}

\begin{thebibliography}{80}
\expandafter\ifx\csname natexlab\endcsname\relax\def\natexlab#1{#1}\fi

\bibitem[{{Ackermann} {et~al.}(2014){Ackermann}, {Ajello}, {Albert},
  {Allafort}, {Baldini}, {Barbiellini}, {Bastieri}, {Bechtol}, {Bellazzini},
  {Bissaldi}, {Bonamente}, {Bottacini}, {Bouvier}, {Brandt}, {Bregeon},
  {Brigida}, {Bruel}, {Buehler}, {Buson}, {Caliandro}, {Cameron}, {Caraveo},
  {Cecchi}, {Charles}, {Chekhtman}, {Chen}, {Chiang}, {Chiaro}, {Ciprini},
  {Claus}, {Cohen-Tanugi}, {Conrad}, {Cutini}, {D'Ammando}, {de Angelis}, {de
  Palma}, {Dermer}, {Desiante}, {Digel}, {Di Venere}, {Silva}, {Drell},
  {Drlica-Wagner}, {Favuzzi}, {Fegan}, {Focke}, {Franckowiak}, {Fukazawa},
  {Funk}, {Fusco}, {Gargano}, {Gasparrini}, {Germani}, {Giglietto}, {Giordano},
  {Giroletti}, {Glanzman}, {Godfrey}, {Grenier}, {Grove}, {Guiriec}, {Hadasch},
  {Hayashida}, {Hays}, {Horan}, {Hughes}, {Inoue}, {Jackson}, {Jogler},
  {J{\'o}hannesson}, {Johnson}, {Kamae}, {Kawano}, {Kn{\"o}dlseder}, {Kuss},
  {Lande}, {Larsson}, {Latronico}, {Lemoine-Goumard}, {Longo}, {Loparco},
  {Lott}, {Lovellette}, {Lubrano}, {Mayer}, {Mazziotta}, {McEnery},
  {Michelson}, {Mizuno}, {Moiseev}, {Monte}, {Monzani}, {Moretti}, {Morselli},
  {Moskalenko}, {Murgia}, {Murphy}, {Nemmen}, {Nuss}, {Ohno}, {Ohsugi},
  {Okumura}, {Omodei}, {Orienti}, {Orlando}, {Ormes}, {Paneque}, {Panetta},
  {Perkins}, {Pesce-Rollins}, {Petrosian}, {Piron}, {Pivato}, {Porter},
  {Rain{\`o}}, {Rando}, {Razzano}, {Reimer}, {Reimer}, {Ritz}, {Schulz},
  {Sgr{\`o}}, {Siskind}, {Spandre}, {Spinelli}, {Takahashi}, {Takeuchi},
  {Tanaka}, {Thayer}, {Thayer}, {Thompson}, {Tibaldo}, {Tinivella}, {Tosti},
  {Troja}, {Tronconi}, {Usher}, {Vandenbroucke}, {Vasileiou}, {Vianello},
  {Vitale}, {Werner}, {Winer}, {Wood}, {Wood}, {Wood}, {Yang}, \& {Fermi LAT
  Collaboration}}]{2014ApJ...787...15A}
{Ackermann}, M., {Ajello}, M., {Albert}, A., {et~al.} 2014, \apj, 787, 15

\bibitem[{Ackermann {et~al.}(2012)Ackermann, Ajello, Allafort, Atwood, Baldini,
  Barbiellini, Bastieri, Bechtol, Bellazzini, Bhat, Blandford, Bonamente,
  Borgland, Bregeon, Briggs, Brigida, Bruel, Buehler, Burgess, Buson,
  Caliandro, Cameron, Casandjian, Cecchi, Charles, Chekhtman, Chiang, Ciprini,
  Claus, Cohen-Tanugi, Connaughton, Conrad, Cutini, Dennis, de~Palma, Dermer,
  Digel, do~Couto~e Silva, Drell, Drlica-Wagner, Dubois, Favuzzi, Fegan,
  Ferrara, Fortin, Fukazawa, Fusco, Gargano, Germani, Giglietto, Giordano,
  Giroletti, Glanzman, Godfrey, Grillo, Grove, Gruber, Guiriec, Hadasch,
  Hayashida, Hays, Horan, Iafrate, Jóhannesson, Johnson, Johnson, Kamae,
  Kippen, Knödlseder, Kuss, Lande, Latronico, Longo, Loparco, Lott,
  Lovellette, Lubrano, Mazziotta, McEnery, Meegan, Mehault, Michelson,
  Mitthumsiri, Monte, Monzani, Morselli, Moskalenko, Murgia, Murphy,
  Naumann-Godo, Nuss, Nymark, Ohno, Ohsugi, Okumura, Omodei, Orlando, Paciesas,
  Panetta, Parent, Pesce-Rollins, Petrosian, Pierbattista, Piron, Pivato, Poon,
  Porter, Preece, Rainò, Rando, Razzano, Razzaque, Reimer, Reimer, Ritz,
  Sbarra, Schwartz, Sgrò, Share, Siskind, Spinelli, Takahashi, Tanaka, Tanaka,
  Thayer, Tibaldo, Tinivella, Tolbert, Tosti, Troja, Uchiyama, Usher,
  Vandenbroucke, Vasileiou, Vianello, Vitale, von Kienlin, Waite, Wilson-Hodge,
  Wood, Wood, \& Yang}]{Ackermann_2012}
Ackermann, M., Ajello, M., Allafort, A., {et~al.} 2012, \apj, 745, 144

\bibitem[{{Ackermann} {et~al.}(2017){Ackermann}, {Allafort}, {Baldini},
  {Barbiellini}, {Bastieri}, {Bellazzini}, {Bissaldi}, {Bonino}, {Bottacini},
  {Bregeon}, {Bruel}, {Buehler}, {Cameron}, {Caragiulo}, {Caraveo},
  {Cavazzuti}, {Cecchi}, {Charles}, {Ciprini}, {Costanza}, {Cutini},
  {D'Ammando}, {de Palma}, {Desiante}, {Digel}, {Di Lalla}, {Di Mauro}, {Di
  Venere}, {Drell}, {Favuzzi}, {Fukazawa}, {Fusco}, {Gargano}, {Giglietto},
  {Giordano}, {Giroletti}, {Grenier}, {Guillemot}, {Guiriec}, {Jogler},
  {J{\'o}hannesson}, {Kashapova}, {Krucker}, {Kuss}, {La Mura}, {Larsson},
  {Latronico}, {Li}, {Liu}, {Longo}, {Loparco}, {Lubrano}, {Magill}, {Maldera},
  {Manfreda}, {Mazziotta}, {Mitthumsiri}, {Mizuno}, {Monzani}, {Morselli},
  {Moskalenko}, {Negro}, {Nuss}, {Ohsugi}, {Omodei}, {Orlando}, {Pal'shin},
  {Paneque}, {Perkins}, {Pesce-Rollins}, {Petrosian}, {Piron}, {Principe},
  {Rain{\`o}}, {Rando}, {Razzano}, {Reimer}, {Rubio da Costa}, {Sgr{\`o}},
  {Simone}, {Siskind}, {Spada}, {Spandre}, {Spinelli}, {Tajima}, {Thayer},
  {Torres}, {Troja}, \& {Vianello}}]{Ackermann_2017}
{Ackermann}, M., {Allafort}, A., {Baldini}, L., {et~al.} 2017, \apj, 835, 219

\bibitem[{{Afanasiev} {et~al.}(2018){Afanasiev}, {Vainio}, {Rouillard},
  {Battarbee}, {Aran}, \& {Zucca}}]{Afa:al-18b}
{Afanasiev}, A., {Vainio}, R., {Rouillard}, A.~P., {et~al.} 2018, \aap, 614, A4

\bibitem[{Ajello {et~al.}(2014)Ajello, Albert, Allafort, Baldini, Barbiellini,
  Bastieri, Bellazzini, Bissaldi, Bonamente, Brandt, Bregeon, Brigida, Bruel,
  Buehler, Buson, Caliandro, Cameron, Caraveo, Cecchi, Charles, Chekhtman,
  Chiang, Chiaro, Ciprini, Claus, Cohen-Tanugi, Cominsky, Conrad, Cutini,
  D'Ammando, de~Palma, Dermer, Desiante, Digel, do~Couto~e Silva, Drell,
  Drlica-Wagner, Favuzzi, Focke, Franckowiak, Fukazawa, Fusco, Gargano,
  Gasparrini, Germani, Giglietto, Giommi, Giordano, Giroletti, Glanzman,
  Godfrey, Grenier, Grove, Guiriec, Hadasch, Hayashida, Hays, Horan, Hou,
  Hughes, Inoue, Jackson, Jogler, Jóhannesson, Johnson, Johnson, Kamae,
  Knödlseder, Kocevski, Kuss, Lande, Larsson, Latronico, Longo, Loparco, Lott,
  Lovellette, Lubrano, Mayer, Mazziotta, McEnery, Michelson, Mizuno, Moiseev,
  Monte, Monzani, Morselli, Moskalenko, Murgia, Murphy, Nakamori, Nemmen, Nuss,
  Ohno, Ohsugi, Omodei, Orienti, Orlando, Ormes, Paneque, Panetta, Perkins,
  Pesce-Rollins, Petrosian, Piron, Pivato, Porter, Rainò, Rando, Razzano,
  Reimer, Reimer, Roth, Schulz, Sgrò, Siskind, Spandre, Spinelli, Takahashi,
  Thayer, Thayer, Thompson, Tibaldo, Tinivella, Tosti, Troja, Usher,
  Vandenbroucke, Vasileiou, Vianello, Vitale, Werner, Winer, Wood, Wood, \&
  Yang}]{Ajello_2014}
Ajello, M., Albert, A., Allafort, A., {et~al.} 2014, \apj, 789, 20

\bibitem[{Ajello {et~al.}(2021)Ajello, Baldini, Bastieri, Bellazzini, Berretta,
  Bissaldi, Blandford, Bonino, Bruel, Buson, Cameron, Caputo, Cavazzuti,
  Cheung, Chiaro, Costantin, Cutini, D'Ammando, de~Palma, Desiante, Lalla,
  Venere, Dirirsa, Fegan, Fukazawa, Funk, Fusco, Gargano, Gasparrini, Giordano,
  Giroletti, Green, Guiriec, Hays, Hewitt, Horan, J{\'{o}}hannesson,
  Kovac'evic', Kuss, Larsson, Latronico, Li, Longo, Lovellette, Lubrano,
  Maldera, Manfreda, Mart{\'{\i}}-Devesa, Mazziotta, Mereu, Michelson, Mizuno,
  Monzani, Morselli, Moskalenko, Negro, Omodei, Orienti, Orlando, Paneque, Pei,
  Persic, Pesce-Rollins, Petrosian, Piron, Porter, Principe, Racusin,
  Rain{\`{o}}, Rando, Rani, Razzano, Razzaque, Reimer, Reimer, Serini,
  Sgr{\`{o}}, Siskind, Spandre, Spinelli, Tak, Troja, Valverde, Wood, \&
  Zaharijas}]{flarecatalog_2021}
Ajello, M., Baldini, L., Bastieri, D., {et~al.} 2021, \apjs, 252, 13

\bibitem[{{Akimov} {et~al.}(1992){Akimov}, {Afanasyev}, {Belaousov},
  {Blokhintsev}, {Volsenskaya}, {Kalinkin}, {Leikov}, {Nesterov}, {Galper},
  {Voronov}, {Zemskov}, {Kirillov-Ugryumov}, {Lutchkov}, {Ozerov}, {Popov},
  {Rudko}, {Runtso}, {Chesnokov}, {Kurnosova}, {Rusakovich}, {Topchiev},
  {Fradkin}, {Chuikin}, {Tugaenko}, {Tian}, {Ishkov}, {Gros}, {Grenier},
  {Barouch}, {Wallin}, {Baser-Bachi}, {Lavigne}, {Olive}, \&
  {Juchniewicz}}]{Aki:al-92}
{Akimov}, V.~V., {Afanasyev}, V.~G., {Belaousov}, A.~S., {et~al.} 1992, Soviet
  Astronomy Letters, 18, 69

\bibitem[{{Aschwanden}(2002)}]{Asc-02}
{Aschwanden}, M.~J. 2002, \ssr, 101, 1

\bibitem[{{Aschwanden} \& {Alexander}(2001)}]{Asc:Alx-01}
{Aschwanden}, M.~J. \& {Alexander}, D. 2001, \solphys, 204, 91

\bibitem[{{Atwood} {et~al.}(2009){Atwood}, {Ackermann}, {Ajello}, {Baldini},
  {Ballet}, {Barbiellini}, {Baring}, {Bastieri}, {Bechtol}, {Bellazzini},
  {Berenji}, {Bhat}, {Bissaldi}, {Blandford}, {Bonamente}, {Bonnell},
  {Borgland}, {Bouvier}, {Bregeon}, {Brigida}, {Bruel}, {Buehler}, {Buson},
  {Caliandro}, {Cameron}, {Caraveo}, {Casandjian}, {Cecchi}, {Charles},
  {Chekhtman}, {Chiang}, {Ciprini}, {Claus}, {Connaughton}, {Conrad}, {Cutini},
  {de Angelis}, {de Palma}, {Dermer}, {Silva}, {Drell}, {Dubois}, {Favuzzi},
  {Fukazawa}, {Fusco}, {Gargano}, {Gehrels}, {Germani}, {Giglietto}, {Giommi},
  {Giordano}, {Giroletti}, {Glanzman}, { Godfrey}, {Granot}, {Grenier},
  {Guiriec}, {Hadasch}, {Hanabata}, {Hughes}, {J{\'o}hannesson}, {Johnson},
  {Kamae}, {Katagiri}, {Kataoka}, {Kerr}, {Kn{\"o}dlseder}, {Kuss}, {Lande},
  {Latronico}, {Lee}, {Longo}, {Loparco}, {Lott}, {Lubrano}, {Mazziotta},
  {McEnery}, {M{\'e}sz{\'a}ros}, {Michelson}, {Mizuno}, {Moiseev}, {Monzani},
  {Morselli}, {Moskalenko}, {Murgia}, {Nakamori}, {Naumann-Godo}, {Nolan},
  {Norris}, {Nuss}, {Ohsugi}, {Okumura}, {Omodei}, {Orlando}, {Paciesas},
  {Pelassa}, {Pesce-Rollins}, {Pierbattista}, {Piron}, {Porter}, {Racusin},
  {Rain{\`o}}, {Razzano}, {Razzaque}, {Reimer}, {Reimer}, { Reyes}, {Roth},
  {Sadrozinski}, {Sgr{\`o}}, {Siskind}, {Smith}, {Sonbas}, {Spandre},
  {Spinelli}, {Stamatikos}, {Strickman}, {Takahashi}, {Tanaka}, {Tanaka},
  {Thayer}, {Thayer}, {Torres}, {Tosti}, {Troja}, {Uehara}, {Usher},
  {Vandenbroucke}, {Vasileiou}, {Vianello}, {Vilchez}, {Vitale}, {von Kienlin},
  {Waite}, {Wang}, {Winer}, {Wood}, {Yamazaki}, {Yang}, {Ziegler}, {Piro}, \&
  {Fermi Collaboration}}]{LATPaper}
{Atwood}, W.~B.{Abdo}, A.~A., {Ackermann}, M., {Ajello}, M., {et~al.} 2009,
  \apj, 697, 1071

\bibitem[{{Barat} {et~al.}(1994){Barat}, {Trottet}, {Vilmer}, {Dezalay},
  {Talon}, {Sunyaev}, {Terekhov}, \& {Kuznetsov}}]{1994ApJ...425L.109B}
{Barat}, C., {Trottet}, G., {Vilmer}, N., {et~al.} 1994, \apjl, 425, L109

\bibitem[{{Benz} {et~al.}(2009){Benz}, {Monstein}, {Meyer}, {Manoharan},
  {Ramesh}, {Altyntsev}, {Lara}, {Paez}, \& {Cho}}]{Ben:al-09}
{Benz}, A.~O., {Monstein}, C., {Meyer}, H., {et~al.} 2009, Earth Moon and
  Planets, 104, 277

\bibitem[{{Boischot} \& {Denisse}(1957)}]{Boi:Den-57}
{Boischot}, A. \& {Denisse}, J.-F. 1957, Acad\'emie des Sciences Paris Comptes
  Rendus, 245, 2194

\bibitem[{{Bougeret} {et~al.}(1995){Bougeret}, {Kaiser}, {Kellogg}, {Manning},
  {Goetz}, {Monson}, {Monge}, {Friel}, {Meetre}, {Perche}, {Sitruk}, \&
  {Hoang}}]{Bou:al-95}
{Bougeret}, J.-L., {Kaiser}, M.~L., {Kellogg}, P.~J., {et~al.} 1995, \ssr, 71,
  231

\bibitem[{Bruno {et~al.}(2023)Bruno, de~Nolfo, Ryan, Richardson, \&
  Dalla}]{Bruno_2023}
Bruno, A., de~Nolfo, G.~A., Ryan, J.~M., Richardson, I.~G., \& Dalla, S. 2023,
  \apj, 953, 187

\bibitem[{{Cai} {et~al.}(2019){Cai}, {Shen}, {Raymond}, {Mei}, {Warmuth},
  {Roussev}, \& {Lin}}]{Cai:al-19}
{Cai}, Q., {Shen}, C., {Raymond}, J.~C., {et~al.} 2019, \mnras, 489, 3183

\bibitem[{{Chen}(2011)}]{Che-11}
{Chen}, P.~F. 2011, Living Reviews in Solar Physics, 8, 1

\bibitem[{{Chupp} {et~al.}(1982){Chupp}, {Forrest}, {Ryan}, {Heslin}, {Reppin},
  {Pinkau}, {Kanbach}, {Rieger}, \& {Share}}]{1982ApJ...263L..95C}
{Chupp}, E.~L., {Forrest}, D.~J., {Ryan}, J.~M., {et~al.} 1982, \apjl, 263, L95

\bibitem[{Chupp \& Ryan(2009)}]{Chupp_2009}
Chupp, E.~L. \& Ryan, J.~M. 2009, Research in Astronomy and Astrophysics, 9, 11

\bibitem[{{Cliver} {et~al.}(1993){Cliver}, {Kahler}, \& {Vestrand}}]{cliv93}
{Cliver}, E.~W., {Kahler}, S.~W., \& {Vestrand}, W.~T. 1993, in International
  Cosmic Ray Conference, Vol.~3, 23rd International Cosmic Ray Conference
  (ICRC23), Volume 3, 91

\bibitem[{{Darnel} {et~al.}(2022){Darnel}, {Seaton}, {Bethge}, {Rachmeler},
  {Jarvis}, {Hill}, {Peck}, {Hughes}, {Shapiro}, {Riley}, {Vasudevan}, {Shing},
  {Koener}, {Edwards}, {Mathur}, \& {Timothy}}]{Darnel:2022}
{Darnel}, J.~M., {Seaton}, D.~B., {Bethge}, C., {et~al.} 2022, Space Weather,
  20, e2022SW003044

\bibitem[{De~Nolfo {et~al.}(2019)De~Nolfo, Bruno, Ryan, Dalla, Giacalone,
  Richardson, Christian, Stochaj, Bazilevskaya, Boezio, Martucci, Mikhailov, \&
  Munini}]{DeNolfo2019}
De~Nolfo, G., Bruno, A., Ryan, J., {et~al.} 2019, \apj, 879, cited By 7

\bibitem[{{Dissauer} {et~al.}(2016){Dissauer}, {Temmer}, {Veronig},
  {Vanninathan}, \& {Magdaleni{\'c}}}]{Dissauer:2016}
{Dissauer}, K., {Temmer}, M., {Veronig}, A.~M., {Vanninathan}, K., \&
  {Magdaleni{\'c}}, J. 2016, \apj, 830, 92

\bibitem[{{Feng} {et~al.}(2013){Feng}, {Wiegelmann}, {Su}, {Inhester}, {Li},
  {Sun}, \& {Gan}}]{2013ApJ...765...37F}
{Feng}, L., {Wiegelmann}, T., {Su}, Y., {et~al.} 2013, \apj, 765, 37

\bibitem[{Forrest {et~al.}(1986)Forrest, Vestrand, Chupp, Rieger, Cooper, \&
  Share}]{FORREST1986115}
Forrest, D., Vestrand, W., Chupp, E., {et~al.} 1986, Advances in Space
  Research, 6, 115

\bibitem[{{Forrest} {et~al.}(1985){Forrest}, {Vestrand}, {Chupp}, {Rieger},
  {Cooper}, \& {Share}}]{1985ICRC....4..146F}
{Forrest}, D.~J., {Vestrand}, W.~T., {Chupp}, E.~L., {et~al.} 1985, in
  International Cosmic Ray Conference, Vol.~4, 19th International Cosmic Ray
  Conference (ICRC19), Volume 4, 146

\bibitem[{{Gopalswamy} {et~al.}(2020){Gopalswamy}, {M{\"a}kel{\"a}}, {Yashiro},
  {Akiyama}, {Xie}, \& {Thakur}}]{2020SoPh..295...18G}
{Gopalswamy}, N., {M{\"a}kel{\"a}}, P., {Yashiro}, S., {et~al.} 2020, \solphys,
  295, 18

\bibitem[{{Gopalswamy} {et~al.}(2018){Gopalswamy}, {M{\"a}kel{\"a}}, {Yashiro},
  {Lara}, {Xie}, {Akiyama}, \& {MacDowall}}]{2018ApJ...868L..19G}
{Gopalswamy}, N., {M{\"a}kel{\"a}}, P., {Yashiro}, S., {et~al.} 2018, \apjl,
  868, L19

\bibitem[{{Gopalswamy} {et~al.}(2012){Gopalswamy}, {Xie}, {Yashiro}, {Akiyama},
  {M{\"a}kel{\"a}}, \& {Usoskin}}]{Gop:al-12}
{Gopalswamy}, N., {Xie}, H., {Yashiro}, S., {et~al.} 2012, \ssr, 171, 23

\bibitem[{Grechnev {et~al.}(2018)Grechnev, Kiselev, Kashapova, Kochanov,
  Zimovets, Uralov, Nizamov, Grigorieva, Golovin, Litvak, Mitrofanov, \&
  Sanin}]{Grechnev_2018}
Grechnev, V.~V., Kiselev, V.~I., Kashapova, L.~K., {et~al.} 2018, \solphys,
  293, 133

\bibitem[{{Hamini} {et~al.}(2021){Hamini}, {Auxepaules}, {Bir{\'e}e},
  {Kenfack}, {Kerdraon}, {Klein}, {Lespagnol}, {Masson}, {Coutouly}, {Fabrice},
  \& {Romagnan}}]{Ham:al-21}
{Hamini}, A., {Auxepaules}, G., {Bir{\'e}e}, L., {et~al.} 2021, \jswsc, 11, 57

\bibitem[{{Heerikhuisen} {et~al.}(2002){Heerikhuisen}, {Litvinenko}, \&
  {Craig}}]{Her:al-02}
{Heerikhuisen}, J., {Litvinenko}, Y.~E., \& {Craig}, I.~J.~D. 2002, \apj, 566,
  512

\bibitem[{{H{\"o}gbom}(1974)}]{1974A&AS...15..417H}
{H{\"o}gbom}, J.~A. 1974, \aaps, 15, 417

\bibitem[{{Hudson}(2018)}]{Hud-18}
{Hudson}, H.~S. 2018, in IAU Symposium, Vol. 335, Space Weather of the
  Heliosphere: Processes and Forecasts, ed. C.~{Foullon} \& O.~E. {Malandraki},
  49--53

\bibitem[{{Hutchinson, A.} {et~al.}(2022){Hutchinson, A.}, {Dalla, S.},
  {Laitinen, T.}, {de Nolfo, G. A.}, {Bruno, A.}, {Ryan, J. M.}, \& {Waterfall,
  C. O. G.}}]{Hutchinson2022}
{Hutchinson, A.}, {Dalla, S.}, {Laitinen, T.}, {et~al.} 2022, A\&A, 658, A23

\bibitem[{{Jin} {et~al.}(2018){Jin}, {Petrosian}, {Liu}, {Nitta}, {Omodei},
  {Rubio da Costa}, {Effenberger}, {Li}, {Pesce-Rollins}, {Allafort}, \&
  {Manchester}}]{Jin2018}
{Jin}, M., {Petrosian}, V., {Liu}, W., {et~al.} 2018, \apj, 867, 122

\bibitem[{{Kanbach} {et~al.}(1993){Kanbach}, {Bertsch}, {Fichtel}, {Hartman},
  {Hunter}, {Kniffen}, {Kwok}, {Lin}, {Mattox}, \&
  {Mayer-Hasselwander}}]{Kanbach1993}
{Kanbach}, G., {Bertsch}, D.~L., {Fichtel}, C.~E., {et~al.} 1993, \aaps, 97,
  349

\bibitem[{{Kane} {et~al.}(1992){Kane}, {McTiernan}, {Loran}, {Fenimore},
  {Klebesadel}, \& {Laros}}]{Kan:al-92}
{Kane}, S.~R., {McTiernan}, J., {Loran}, J., {et~al.} 1992, \apj, 390, 687

\bibitem[{{Kerdraon} \& {Delouis}(1997)}]{Ker:Del-97}
{Kerdraon}, A. \& {Delouis}, J.-M. 1997, in Lecture Notes in Physics, Vol. 483,
  Coronal Physics from Radio and Space Observations, ed. G.~{Trottet} (Berlin,
  Heidelberg, New York...: Springer), 192--201

\bibitem[{{Klassen} {et~al.}(2000){Klassen}, {Aurass}, {Mann}, \&
  {Thompson}}]{Kla:al-00}
{Klassen}, A., {Aurass}, H., {Mann}, G., \& {Thompson}, B.~J. 2000, \aaps, 141,
  357

\bibitem[{{Klein} {et~al.}(1999){Klein}, {Chupp}, {Trottet}, {Magun}, {Dunphy},
  {Rieger}, \& {Urpo}}]{Kle:al-99a}
{Klein}, K.-L., {Chupp}, E.~L., {Trottet}, G., {et~al.} 1999, \aap, 348, 271

\bibitem[{{Klein} {et~al.}(2022){Klein}, {Musset}, {Vilmer}, {Briand},
  {Krucker}, {Battaglia}, {Dresing}, {Palmroos}, \& {Gary}}]{Kle:al-22}
{Klein}, K.-L., {Musset}, S., {Vilmer}, N., {et~al.} 2022, \aap, 663, A173

\bibitem[{{Klein} {et~al.}(2018){Klein}, {Tziotziou}, {Zucca}, {Valtonen},
  {Vilmer}, {Malandraki}, {Hamadache}, {Heber}, \& {Kiener}}]{Kle:al-18}
{Klein}, K.-L., {Tziotziou}, K., {Zucca}, P., {et~al.} 2018, in Astrophysics
  and Space Science Library, Vol. 444, Solar Particle Radiation Storms
  Forecasting and Analysis, ed. O.~E. {Malandraki} \& N.~B. {Crosby}, 133--155

\bibitem[{{Krucker} {et~al.}(2020){Krucker}, {Hurford}, {Grimm}, {K{\"o}gl},
  {Gr{\"o}belbauer}, {Etesi}, {Casadei}, {Csillaghy}, {Benz}, {Arnold},
  {Molendini}, {Orleanski}, {Schori}, {Xiao}, {Kuhar}, {Hochmuth}, {Felix},
  {Schramka}, {Marcin}, {Kobler}, {Iseli}, {Dreier}, {Wiehl}, {Kleint},
  {Battaglia}, {Lastufka}, {Sathiapal}, {Lapadula}, {Bednarzik}, {Birrer},
  {Stutz}, {Wild}, {Marone}, {Skup}, {Cichocki}, {Ber}, {Rutkowski}, {Bujwan},
  {Juchnikowski}, {Winkler}, {Darmetko}, {Michalska}, {Seweryn}, {Bia{\l}ek},
  {Osica}, {Sylwester}, {Kowalinski}, {{\'S}cis{\l}owski}, {Siarkowski},
  {St{\k{e}}{\'s}licki}, {Mrozek}, {Podg{\'o}rski}, {Meuris}, {Limousin},
  {Gevin}, {Le Mer}, {Brun}, {Strugarek}, {Vilmer}, {Musset}, {Maksimovi{\'c}},
  {F{\'a}rn{\'\i}k}, {Koz{\'a}{\v{c}}ek}, {Ka{\v{s}}parov{\'a}}, {Mann},
  {{\"O}nel}, {Warmuth}, {Rendtel}, {Anderson}, {Bauer}, {Dionies}, {Paschke},
  {Pl{\"u}schke}, {Woche}, {Schuller}, {Veronig}, {Dickson}, {Gallagher},
  {Maloney}, {Bloomfield}, {Piana}, {Massone}, {Benvenuto}, {Massa},
  {Schwartz}, {Dennis}, {van Beek}, {Rodr{\'\i}guez-Pacheco}, \&
  {Lin}}]{2020A&A...642A..15K}
{Krucker}, S., {Hurford}, G.~J., {Grimm}, O., {et~al.} 2020, \aap, 642, A15

\bibitem[{Krucker {et~al.}(2011)Krucker, Kontar, Christe, Glesener, \&
  Lin}]{Krucker_2011}
Krucker, S., Kontar, E.~P., Christe, S., Glesener, L., \& Lin, R.~P. 2011,
  \apj, 742, 82

\bibitem[{Lecacheux(2000)}]{Lec:al-00}
Lecacheux, A. 2000, Washington DC American Geophysical Union Geophysical
  Monograph Series, 119

\bibitem[{{Lemen} {et~al.}(2012){Lemen}, {Title}, {Akin}, {Boerner}, {Chou},
  {Drake}, {Duncan}, {Edwards}, {Friedlaender}, {Heyman}, {Hurlburt}, {Katz},
  {Kushner}, {Levay}, {Lindgren}, {Mathur}, {McFeaters}, {Mitchell}, {Rehse},
  {Schrijver}, {Springer}, {Stern}, {Tarbell}, {Wuelser}, {Wolfson}, {Yanari},
  {Bookbinder}, {Cheimets}, {Caldwell}, {Deluca}, {Gates}, {Golub}, {Park},
  {Podgorski}, {Bush}, {Scherrer}, {Gummin}, {Smith}, {Auker}, {Jerram},
  {Pool}, {Soufli}, {Windt}, {Beardsley}, {Clapp}, {Lang}, \&
  {Waltham}}]{Lem:al-12}
{Lemen}, J.~R., {Title}, A.~M., {Akin}, D.~J., {et~al.} 2012, \solphys, 275, 17

\bibitem[{{Litvinenko} \& {Somov}(1995)}]{Lit:Som-95}
{Litvinenko}, Y.~E. \& {Somov}, B.~V. 1995, \solphys, 158, 317

\bibitem[{{Macrae} {et~al.}(2018){Macrae}, {Zharkov}, {Zharkova}, {Druett},
  {Matthews}, \& {Kawate}}]{SunQuakes}
{Macrae}, C., {Zharkov}, S., {Zharkova}, V., {et~al.} 2018, A\&A, 619, A65

\bibitem[{{Mandzhavidze} \& {Ramaty}(1992)}]{1992ApJ...396L.111M}
{Mandzhavidze}, N. \& {Ramaty}, R. 1992, \apjl, 396, L111

\bibitem[{{Mann} {et~al.}(2022){Mann}, {Vocks}, {Warmuth}, {Magdalenic},
  {Bisi}, {Carley}, {Dabrowski}, {Gallagher}, {Krankowski}, {Matyjasiak},
  {Rotkaehl}, \& {Zucca}}]{Man:al-22}
{Mann}, G., {Vocks}, C., {Warmuth}, A., {et~al.} 2022, \aap, 660, A71

\bibitem[{{Masson} {et~al.}(2009){Masson}, {Klein}, {B{\"u}tikofer},
  {Fl{\"u}ckiger}, {Kurt}, {Yushkov}, \& {Krucker}}]{Msn:al-09}
{Masson}, S., {Klein}, K.-L., {B{\"u}tikofer}, R., {et~al.} 2009, \solphys,
  257, 305

\bibitem[{Mattox {et~al.}(1996)Mattox, Bertsch, Chiang, Dingus, Digel,
  Esposito, Fierro, Hartman, Hunter, Kanbach, Kniffen, Lin, Macomb,
  Mayer-Hasselwander, Michelson, von Montigny, Mukherjee, Nolan, Ramanamurthy,
  Schneid, Sreekumar, Thompson, \& Willis}]{Mattox:96}
Mattox, J.~R., Bertsch, D.~L., Chiang, J., {et~al.} 1996, \apj, 461, 396

\bibitem[{Meegan {et~al.}(2009)Meegan, Lichti, Bhat, Bissaldi, Briggs,
  Connaughton, Diehl, Fishman, Greiner, Hoover, van~der Horst, von Kienlin,
  Kippen, Kouveliotou, McBreen, Paciesas, Preece, Steinle, Wallace, Wilson, \&
  Wilson-Hodge}]{Meegan_2009}
Meegan, C., Lichti, G., Bhat, P.~N., {et~al.} 2009, \apj, 702, 791

\bibitem[{{M{\"u}ller} {et~al.}(2020){M{\"u}ller}, {St. Cyr}, {Zouganelis},
  {Gilbert}, {Marsden}, {Nieves-Chinchilla}, {Antonucci}, {Auch{\`e}re},
  {Berghmans}, {Horbury}, {Howard}, {Krucker}, {Maksimovic}, {Owen}, {Rochus},
  {Rodriguez-Pacheco}, {Romoli}, {Solanki}, {Bruno}, {Carlsson}, {Fludra},
  {Harra}, {Hassler}, {Livi}, {Louarn}, {Peter}, {Sch{\"u}hle}, {Teriaca}, {del
  Toro Iniesta}, {Wimmer-Schweingruber}, {Marsch}, {Velli}, {De Groof},
  {Walsh}, \& {Williams}}]{2020A&A...642A...1M}
{M{\"u}ller}, D., {St. Cyr}, O.~C., {Zouganelis}, I., {et~al.} 2020, \aap, 642,
  A1

\bibitem[{{Murphy} {et~al.}(1987){Murphy}, {Dermer}, \& {Ramaty}}]{murp87}
{Murphy}, R.~J., {Dermer}, C.~D., \& {Ramaty}, R. 1987, \apjs, 63, 721

\bibitem[{Pesce-Rollins {et~al.}(2022)Pesce-Rollins, Omodei, Krucker, Lalla,
  Wang, Battaglia, Warmuth, Veronig, \& Baldini}]{Pesce-Rollins_2022}
Pesce-Rollins, M., Omodei, N., Krucker, S., {et~al.} 2022, \apj, 929, 172

\bibitem[{{Pesnell} {et~al.}(2012){Pesnell}, {Thompson}, \&
  {Chamberlin}}]{Pesnell:2012}
{Pesnell}, W.~D., {Thompson}, B.~J., \& {Chamberlin}, P.~C. 2012, \solphys,
  275, 3

\bibitem[{{Plotnikov} {et~al.}(2017){Plotnikov}, {Rouillard}, \&
  {Share}}]{Plotnikov_2017}
{Plotnikov}, I., {Rouillard}, A.~P., \& {Share}, G.~H. 2017, A\&A, 608, A43

\bibitem[{{Rank, G.} {et~al.}(2001){Rank, G.}, {Ryan, J.}, {Debrunner, H.},
  {McConnell, M.}, \& {Sch\"onfelder, V.}}]{rank2001}
{Rank, G.}, {Ryan, J.}, {Debrunner, H.}, {McConnell, M.}, \& {Sch\"onfelder,
  V.} 2001, A\&A, 378, 1046

\bibitem[{{Reames}(2009)}]{Rea-09}
{Reames}, D.~V. 2009, \apj, 693, 812

\bibitem[{{Rieger} {et~al.}(1999){Rieger}, {Treumann}, \&
  {Karlick{\'y}}}]{Rgr:al-99}
{Rieger}, E., {Treumann}, R.~A., \& {Karlick{\'y}}, M. 1999, \solphys, 187, 59

\bibitem[{{Rochus, P.} {et~al.}(2020){Rochus, P.}, {Auch\`ere, F.}, {Berghmans,
  D.}, {Harra, L.}, {Schmutz, W.}, {Sch\"uhle, U.}, {Addison, P.},
  {Appourchaux, T.}, {Aznar Cuadrado, R.}, {Baker, D.}, {Barbay, J.}, {Bates,
  D.}, {BenMoussa, A.}, {Bergmann, M.}, {Beurthe, C.}, {Borgo, B.}, {Bonte,
  K.}, {Bouzit, M.}, {Bradley, L.}, {B\"uchel, V.}, {Buchlin, E.}, {B\"uchner,
  J.}, {Cab\'e, F.}, {Cadiergues, L.}, {Chaigneau, M.}, {Chares, B.}, {Choque
  Cortez, C.}, {Coker, P.}, {Condamin, M.}, {Coumar, S.}, {Curdt, W.}, {Cutler,
  J.}, {Davies, D.}, {Davison, G.}, {Defise, J.-M.}, {Del Zanna, G.},
  {Delmotte, F.}, {Delouille, V.}, {Dolla, L.}, {Dumesnil, C.}, {D\"urig, F.},
  {Enge, R.}, {Fran\c{c}ois, S.}, {Fourmond, J.-J.}, {Gillis, J.-M.},
  {Giordanengo, B.}, {Gissot, S.}, {Green, L. M.}, {Guerreiro, N.}, {Guilbaud,
  A.}, {Gyo, M.}, {Haberreiter, M.}, {Hafiz, A.}, {Hailey, M.}, {Halain,
  J.-P.}, {Hansotte, J.}, {Hecquet, C.}, {Heerlein, K.}, {Hellin, M.-L.},
  {Hemsley, S.}, {Hermans, A.}, {Hervier, V.}, {Hochedez, J.-F.}, {Houbrechts,
  Y.}, {Ihsan, K.}, {Jacques, L.}, {J\'er\^ome, A.}, {Jones, J.}, {Kahle, M.},
  {Kennedy, T.}, {Klaproth, M.}, {Kolleck, M.}, {Koller, S.}, {Kotsialos, E.},
  {Kraaikamp, E.}, {Langer, P.}, {Lawrenson, A.}, {Le Clech\'{}, J.-C.},
  {Lenaerts, C.}, {Liebecq, S.}, {Linder, D.}, {Long, D. M.}, {Mampaey, B.},
  {Markiewicz-Innes, D.}, {Marquet, B.}, {Marsch, E.}, {Matthews, S.}, {Mazy,
  E.}, {Mazzoli, A.}, {Meining, S.}, {Meltchakov, E.}, {Mercier, R.}, {Meyer,
  S.}, {Monecke, M.}, {Monfort, F.}, {Morinaud, G.}, {Moron, F.}, {Mountney,
  L.}, {M\"uller, R.}, {Nicula, B.}, {Parenti, S.}, {Peter, H.}, {Pfiffner,
  D.}, {Philippon, A.}, {Phillips, I.}, {Plesseria, J.-Y.}, {Pylyser, E.},
  {Rabecki, F.}, {Ravet-Krill, M.-F.}, {Rebellato, J.}, {Renotte, E.},
  {Rodriguez, L.}, {Roose, S.}, {Rosin, J.}, {Rossi, L.}, {Roth, P.},
  {Rouesnel, F.}, {Roulliay, M.}, {Rousseau, A.}, {Ruane, K.}, {Scanlan, J.},
  {Schlatter, P.}, {Seaton, D. B.}, {Silliman, K.}, {Smit, S.}, {Smith, P. J.},
  {Solanki, S. K.}, {Spescha, M.}, {Spencer, A.}, {Stegen, K.}, {Stockman, Y.},
  {Szwec, N.}, {Tamiatto, C.}, {Tandy, J.}, {Teriaca, L.}, {Theobald, C.},
  {Tychon, I.}, {van Driel-Gesztelyi, L.}, {Verbeeck, C.}, {Vial, J.-C.},
  {Werner, S.}, {West, M. J.}, {Westwood, D.}, {Wiegelmann, T.}, {Willis, G.},
  {Winter, B.}, {Zerr, A.}, {Zhang, X.}, \& {Zhukov, A. N.}}]{SO_EUIpaper}
{Rochus, P.}, {Auch\`ere, F.}, {Berghmans, D.}, {et~al.} 2020, A\&A, 642, A8

\bibitem[{{Ryan}(2000)}]{ryan00}
{Ryan}, J.~M. 2000, \ssr, 93, 581

\bibitem[{{Ryan} \& {Lee}(1991)}]{ryanlee91}
{Ryan}, J.~M. \& {Lee}, M.~A. 1991, \apj, 368, 316

\bibitem[{Share {et~al.}(2018)Share, Murphy, White, Tolbert, Dennis, Schwartz,
  Smart, \& Shea}]{Share_2018}
Share, G.~H., Murphy, R.~J., White, S.~M., {et~al.} 2018, \apj, 869, 182

\bibitem[{Shih {et~al.}(2009)Shih, Lin, \& Smith}]{Shih_2009}
Shih, A.~Y., Lin, R.~P., \& Smith, D.~M. 2009, \apj, 698, L152

\bibitem[{{Sinclair Reid} \& {Ratcliffe}(2014)}]{ScR-Rat-14}
{Sinclair Reid}, H.~A. \& {Ratcliffe}, H. 2014, Research in Astronomy and
  Astrophysics, 14, 773

\bibitem[{{Tan} {et~al.}(2016){Tan}, {M{\'e}sz{\'a}rosov{\'a}}, {Karlick{\'y}},
  {Huang}, \& {Tan}}]{TBL:al-16}
{Tan}, B., {M{\'e}sz{\'a}rosov{\'a}}, H., {Karlick{\'y}}, M., {Huang}, G., \&
  {Tan}, C. 2016, \apj, 819, 42

\bibitem[{{Trottet} {et~al.}(1998){Trottet}, {Vilmer}, {Barat}, {Benz},
  {Magun}, {Kuznetsov}, {Sunyaev}, \& {Terekhov}}]{Tro:al-98}
{Trottet}, G., {Vilmer}, N., {Barat}, C., {et~al.} 1998, \aap, 334, 1099

\bibitem[{{Vestrand} \& {Forrest}(1993)}]{1993ApJ...409L..69V}
{Vestrand}, W.~T. \& {Forrest}, D.~J. 1993, \apjl, 409, L69

\bibitem[{{Vilmer} {et~al.}(1982){Vilmer}, {Kane}, \&
  {Trottet}}]{1982A&A...108..306V}
{Vilmer}, N., {Kane}, S.~R., \& {Trottet}, G. 1982, \aap, 108, 306

\bibitem[{{Vilmer} {et~al.}(2003){Vilmer}, {Pick}, {Schwenn}, {Ballatore}, \&
  {Villain}}]{Vil:al-03}
{Vilmer}, N., {Pick}, M., {Schwenn}, R., {Ballatore}, P., \& {Villain}, J.~P.
  2003, Annales Geophysicae, 21, 847

\bibitem[{{Vilmer} {et~al.}(1999){Vilmer}, {Trottet}, {Barat}, {Schwartz},
  {Enome}, {Kuznetsov}, {Sunyaev}, \& {Terekhov}}]{1999A&A...342..575V}
{Vilmer}, N., {Trottet}, G., {Barat}, C., {et~al.} 1999, \aap, 342, 575

\bibitem[{{Warren} {et~al.}(2018){Warren}, {Brooks}, {Ugarte-Urra}, {Reep},
  {Crump}, \& {Doschek}}]{Wrr:al-18}
{Warren}, H.~P., {Brooks}, D.~H., {Ugarte-Urra}, I., {et~al.} 2018, \apj, 854,
  122

\bibitem[{{Wild} {et~al.}(1963){Wild}, {Smerd}, \& {Weiss}}]{Wil:al-63}
{Wild}, J.~P., {Smerd}, S.~F., \& {Weiss}, A.~A. 1963, \araa, 1, 291

\bibitem[{{Wu} {et~al.}(2021){Wu}, {Rouillard}, {Kouloumvakos}, {Vainio},
  {Afanasiev}, {Plotnikov}, {Murphy}, {Mann}, \& {Warmuth}}]{wu2021}
{Wu}, Y., {Rouillard}, A.~P., {Kouloumvakos}, A., {et~al.} 2021, \apj, 909, 163

\bibitem[{Xu {et~al.}(2014)Xu, Jing, Wang, \& Wang}]{Xu_2014}
Xu, Y., Jing, J., Wang, S., \& Wang, H. 2014, \apj, 787, 7

\bibitem[{{Yashiro} {et~al.}(2004){Yashiro}, {Gopalswamy}, {Michalek},
  {St.~Cyr}, {Plunkett}, {Rich}, \& {Howard}}]{Yas:al-04}
{Yashiro}, S., {Gopalswamy}, N., {Michalek}, G., {et~al.} 2004, \jgr, 109,
  A07105

\bibitem[{{Zhang} {et~al.}(2023){Zhang}, {Reid}, {Krupar}, {Zucca},
  {Dabrowski}, \& {Krankowski}}]{ZhaJ:al_23}
{Zhang}, J., {Reid}, H. A.~S., {Krupar}, V., {et~al.} 2023, \solphys, 298, 7

\end{thebibliography}

\end{document}